\documentclass{aa}              
\usepackage[english]{babel}
\usepackage[utf8]{inputenc}
\usepackage{amsmath,amsfonts,amssymb,enumerate}
\usepackage{txfonts,graphics,graphicx,lscape,longtable,subcaption,caption,hyperref}
\usepackage{rotating}

\newcommand*\xbar[1]{{\hbox{\vbox{\hrule height 0.5pt \kern0.5ex \hbox{\kern-0.1em \ensuremath{#1} \kern-0.1em}}}}}

\usepackage{natbib}
        \bibpunct{(}{)}{;}{a}{}{,} 

\def\apjl{{Astrophys. J. Lett.} }

\def\aap{{Astron. Astrophys.} }

\def\apjs {{Astrophys. J. Suppl.}}

\begin{document}

\title{On the fragmentation of filaments in a molecular cloud simulation}
\author{R.-A.~Chira\inst{\ref{mpia},\ref{eso}} \and
  J.~Kainulainen\inst{\ref{mpia}} \and
  J.~C.~Ib\'a\~{n}ez-Mej\'{\i}a\inst{\ref{koeln},\ref{mpe}} \and
  Th.~Henning\inst{\ref{mpia}} \and 
  M.-M.~Mac~Low\inst{\ref{amnh},\ref{ita}}}
\institute{
        Max-Planck-Institut f\"ur Astronomie, K\"onigstuhl 17, 69117 Heidelberg, Germany\\ \email{rox.chira@gmail.com}\label{mpia}
        \and European Southern Observatory, Karl-Schwarzschild-Str.\
        2, 85748 Garching bei M\"unchen, Germany\label{eso}
        \and I.\ Physikalisches Institut, Universit\"at zu K\"oln,
        Z\"ulpicher Straße 77, 50937 K\"oln, Germany\label{koeln}
        \and Max-Planck-Institut f\"ur Extraterrestrische Physik,
          Giessenbachstrasse 1, 85748 Garching, Germany\label{mpe}
        \and Dept.\ of Astrophysics, American Museum of Natural History, 79th St.\ at Central Park West, New York, NY 10024, USA\label{amnh}
        \and Zentrum f\"ur Astronomie, Institut f\"ur Theoretische
        Astrophysik, Universit\"at Heidelberg, Albert-Ueberle-Str.\ 2, 69120 Heidelberg, Germany\label{ita}
}

\date{\today}

\abstract
        { 
        The fragmentation of filaments in molecular clouds has attracted a lot of attention recently as there seems to be a close relation between the evolution of filaments and star formation. The study of the fragmentation process has been motivated by simple analytical models. However, only a few comprehensive studies have analysed the evolution of filaments using numerical simulations where the filaments form self-consistently as part of large-scale molecular cloud evolution.
    }
        { 
        We address the early evolution of parsec-scale filaments that form within individual clouds. In particular, we focus on three questions: How do the line masses of filaments evolve? How and when do the filaments fragment? How does the fragmentation relate to the line masses of the filaments?
    } 
        { 
                We examine three simulated molecular clouds formed in kiloparsec-scale numerical simulations performed with the FLASH adaptive mesh refinement magnetohydrodynamic code. The simulations model a self-gravitating, magnetised, stratified, supernova-driven interstellar medium, including photoelectric heating and radiative cooling. We follow the evolution of the clouds for 6~Myr from the time self-gravity starts to act. We identify filaments using the \texttt{DisPerSe} algorithm, and compare the results to other filament-finding algorithms. We determine the properties of the identified filaments and compare them with the predictions of analytic filament stability models.
    }
        { 
        The average line masses of the identified filaments, as well as the fraction of mass in filamentary structures, increases fairly continuously after the onset of self-gravity. The filaments show fragmentation starting relatively early: the first fragments appear when the line masses lie well below the critical line mass of Ostriker's isolated hydrostatic equilibrium solution ($\sim$16~M$_\odot$~pc$^{-1}$), commonly used as a fragmentation criterion. The average line masses of filaments identified in three-dimensional volume density cubes increases far more quickly than those identified in two-dimensional column density maps.
    }
        { 
        Our results suggest that hydrostatic or dynamic compression from the surrounding cloud has a significant impact on the early dynamical evolution of filaments. A simple model of an isolated, isothermal cylinder may not provide a good approach for fragmentation analysis. Caution must be exercised in interpreting distributions of properties of filaments identified in column density maps, especially in the case of low-mass filaments. Comparing or combining results from studies that use different filament finding techniques is strongly discouraged.
    }
        

        \maketitle 

\section{Introduction}\label{intro}

Filamentary structures in the interstellar medium (ISM) have been known and investigated for many years.
Much effort has been invested in studying their morphology \citep{Barnard1927,Schneider1979}, properties \citep{Schmalzl2010,Hacar2013,Andre2014,Smith2014b,Kainulainen2017}, distribution within the Milky Way \citep{Molinari2010,Ragan2014,Abreu-Vicente2016,Li2016}, and formation \citep{Rivera2015,Federrath2016,Smith2016}.
More recently, it has been argued that filaments represent a crucial phase in the earliest stages of star formation, based on observational findings showing that pre-stellar cores and young stellar clusters are preferentially located within filaments or at intersections between them \citep{Schmalzl2010,Myers2011,Schneider2012,Koenyves2015}.
This raises the question of what exactly the role of filaments is in controlling star formation \citep[e.g.][]{Zhang2009,Gomez2014,Pineda2015,Henshaw2016b}.

The usual model for filaments describes them as hydrostatic cylinders that fragment due to linear perturbations \citep{Ostriker1964a,Larson1985,Padoan1999}.
However, some studies \citep[e.g. by][]{Lee1999,Hartmann2002,Henshaw2016b,Clarke2017,Gritschneder2017} demonstrate that other environmental conditions, such as turbulence, accretion, or magnetic fields, can introduce additional fragmentation modes.
Therefore, the question is whether a quasi-static description of the evolution of filaments is justified.
One way to test this is to use numerical simulations that form filaments self-consistently and compare the properties of the resulting structures with those predicted by the simple model.
In doing so, one can examine how each force, and combination of different forces, affects the evolution of filaments, and whether a quasi-static model can capture the essential physics involved in the process of fragmentation.

Since the forces acting on filaments are not directly observable, one needs to analyse the properties of the filaments and fragments as observational diagnostics.
The width of filaments \citep[among others,][]{Arzoumanian2011,Hacar2011,Juvela2012a,Malinen2012,Palmeirim2013,Smith2014b} and the spacing and distribution of fragments \citep[e.g.][]{Jackson2010,Kainulainen2013c,Wang2014,Beuther2015,Ragan2015,Contreras2016,Henshaw2016b} are widely-studied quantities that can be used.
However, these quantities are highly degenerate in this context. 
For example, \citet{Federrath2016} has demonstrated that, as long as self-gravity is not acting alone, the widths of the radial profiles of filaments are approximately constant over wide ranges of central density and global star formation efficiencies.
Furthermore, since filaments are generally short-lived substructures within globally fragmenting molecular clouds and normally neither quiescent nor isolated, external gravitational potentials or pressures can introduce additional perturbations that either separate or compress the cores.

Another interesting parameter is the mass per unit length, or line mass, of a filament, which is argued to determine the filament's stability in the quasi-static model \citep{Ostriker1964a,Ostriker1964b,Nagasawa1987,Inutsuka1992,Fiege2000a,Fischera2012a}.
Similar to the Jeans analysis for spheres, there is a critical line mass that marks the transition between states of equilibrium and gravitational collapse. 
In principle, the line mass is an easy-to-measure quantity since it only requires the length and the enclosed mass of the respective filament.
However, there are many effects that need to be considered when measuring the line mass from observational data, such as inclination, optical depth, uncertainties in distance estimations, and overlap effects \citep{Ballesteros2002,ZamoraAviles2017}.
Furthermore, and especially in the context of large filaments on galactic scales \citep{Ragan2014,Zucker2015,Abreu-Vicente2016}, it is not completely clear how mostly two-dimensional (2D) observational characteristics reflect the three-dimensional (3D) nature of the acting forces. 

The questions we address in this paper are:
How do filaments evolve and fragment in numerical simulations? 
Is the fragmentation picture seen in simulations in agreement with the quasi-static analytic framework of gravitational fragmentation?
To address these questions, we analyse a set of model clouds from 3D FLASH adaptive mesh refinement~(AMR) simulations. 
We introduce the simulations and the analysis methods in Sect.~\ref{methods}.
We tested several filament finder codes and compared their performances in identifying filaments from simulations in Appendix~\ref{a_filfinder}.
The results of our analysis are outlined and discussed in Sect.~\ref{results} and summarised in Sect.~\ref{conclusions}.

\section{Methods}\label{methods}

\subsection{Cloud models}\label{methods_cloudmodel}

The filamentary structures we analyse in this paper come from numerical simulations of dense cloud formation by \citet{IbanezMejia2016}.
The simulations have been performed with the 3D magnetohydrodynamics~(MHD), AMR FLASH code \citep{Fryxell2000}, and are of a $1\times1\times 40$~kpc$^3$ vertical box of the ISM in a disk galaxy, with the galactic midplane located in the middle of the box.
Turbulence is driven by the injection of discrete, thermal supernova~(SN) explosions. 
SN rates are normalized to the galactic SN rates \citep{Tammann1994TheRate}, with 6.58 and 27.4~Myr$^{-1}$~kpc$^{-2}$ for Type~Ia and core-collapse SNe, respectively.
The positioning of SN explosions is random in the horizontal direction and exponentially decaying in the vertical direction with scale heights of 90 and 325~pc for type~Ia and core-collapse SNe.
SN clustering is also taken into account by assuming three fifths of the core-collapse population are correlated in space and time.
Magnetic fields are included in the simulation, with an initial field strength of 5~$\mu$G at the midplane that exponentially decays in the vertical direction. 
The fields are initially uniform and oriented in the horizontal direction. 
They are allowed to evolve self-consistently with the simulation.
The simulations also include distributed photoelectric heating and radiative cooling.
The photoelectric heating of dust grains \citep{Bakes1994TheHydrocarbons} is implemented with a background FUV intensity of $G_{0}=1.7$ with a vertical scale height of 300~pc.
Radiative cooling rates are appropriate for a solar metallicity, optically thin gas, with a constant ionization fraction of $\chi_{e}=0.01$ for temperatures below $2\times 10^{4}$~K \citep{Dalgarno1972HeatingRegions}, and resonant line cooling for collisionally ionized gas at higher temperatures \citep{Sutherland1993CoolingPlasmas}.

The simulation initially includes only a static galactic gravitational potential, accounting for a stellar disk and a spherical dark matter halo.
The potential is uniform in the horizontal direction; in the vertical direction, for altitudes $z \leq 8$~kpc, it follows a parametrized model of the mass distribution of the Milky Way \citep{Dehnen1998MassWay}, and at higher altitudes, it smoothly transforms into the outer halo profile described by \citet{Navarro1996TheHalos} with a scale length of 20~kpc \citep{Hill2012}.
The simulation runs for $\sim$250~Myr without gas self-gravity in order to develop a multiphase, turbulent ISM, where dense structures form self-consistently in turbulent, convergent flows.
Self-gravity is then switched on to follow the formation and evolution of clouds over the next 6~Myr.

\begin{figure*}
        \centering
        \begin{subfigure}{0.32\textwidth}
        \includegraphics[width=\textwidth]{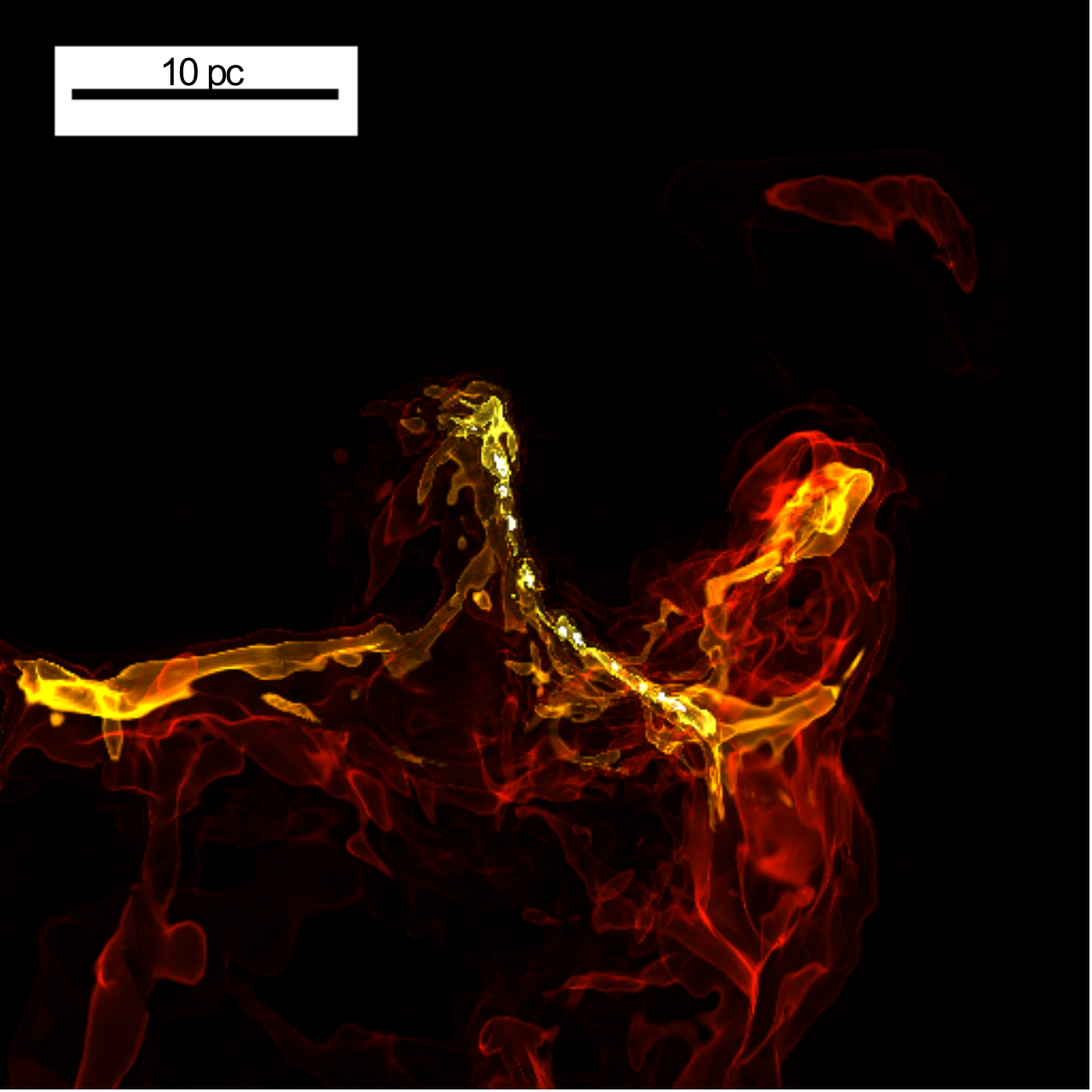}
        \caption{\texttt{M3}, contour colour table}
        \label{pic_methods_m3e3_0040_3d_v1}
    \end{subfigure}
        \begin{subfigure}{0.32\textwidth}
        \includegraphics[width=\textwidth]{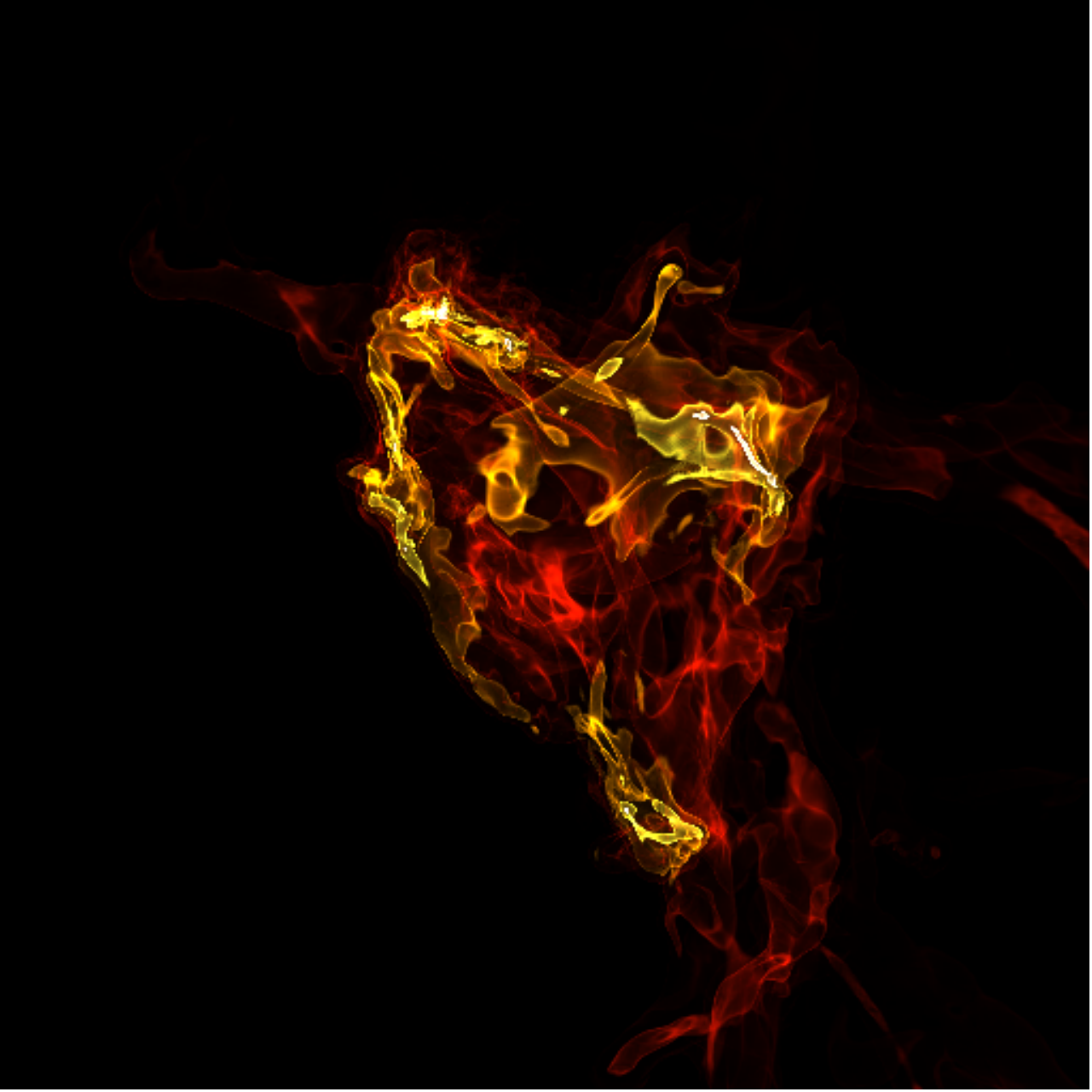}
        \caption{\texttt{M4}, contour colour table}
        \label{pic_methods_m4e3_0040_3d_v1}
    \end{subfigure}
        \begin{subfigure}{0.32\textwidth}
        \includegraphics[width=\textwidth]{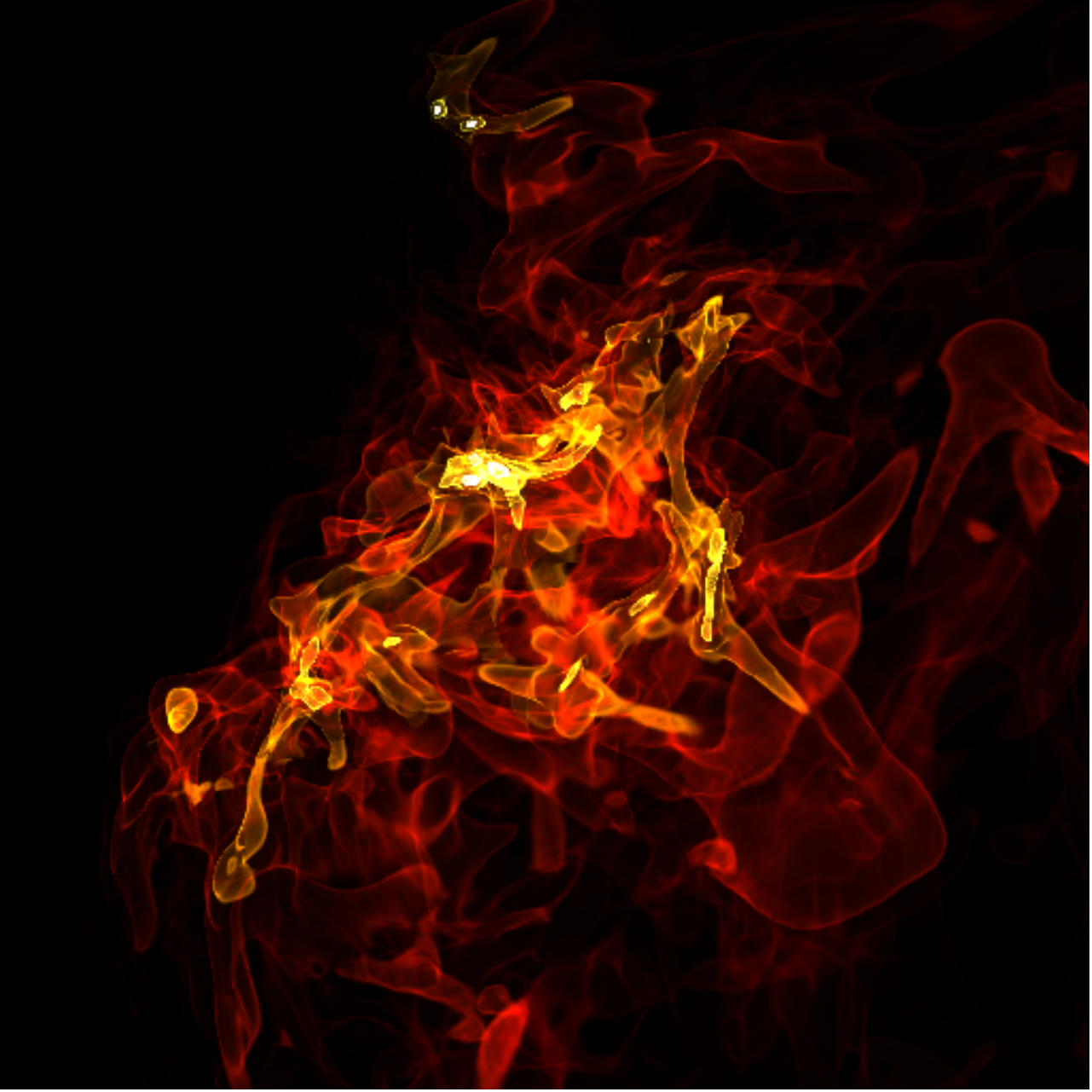}
        \caption{\texttt{M8}, contour colour table}
        \label{pic_methods_m8e3_0040_3d_v1}
    \end{subfigure}
        
        \begin{subfigure}{0.32\textwidth}
        \includegraphics[width=\textwidth]{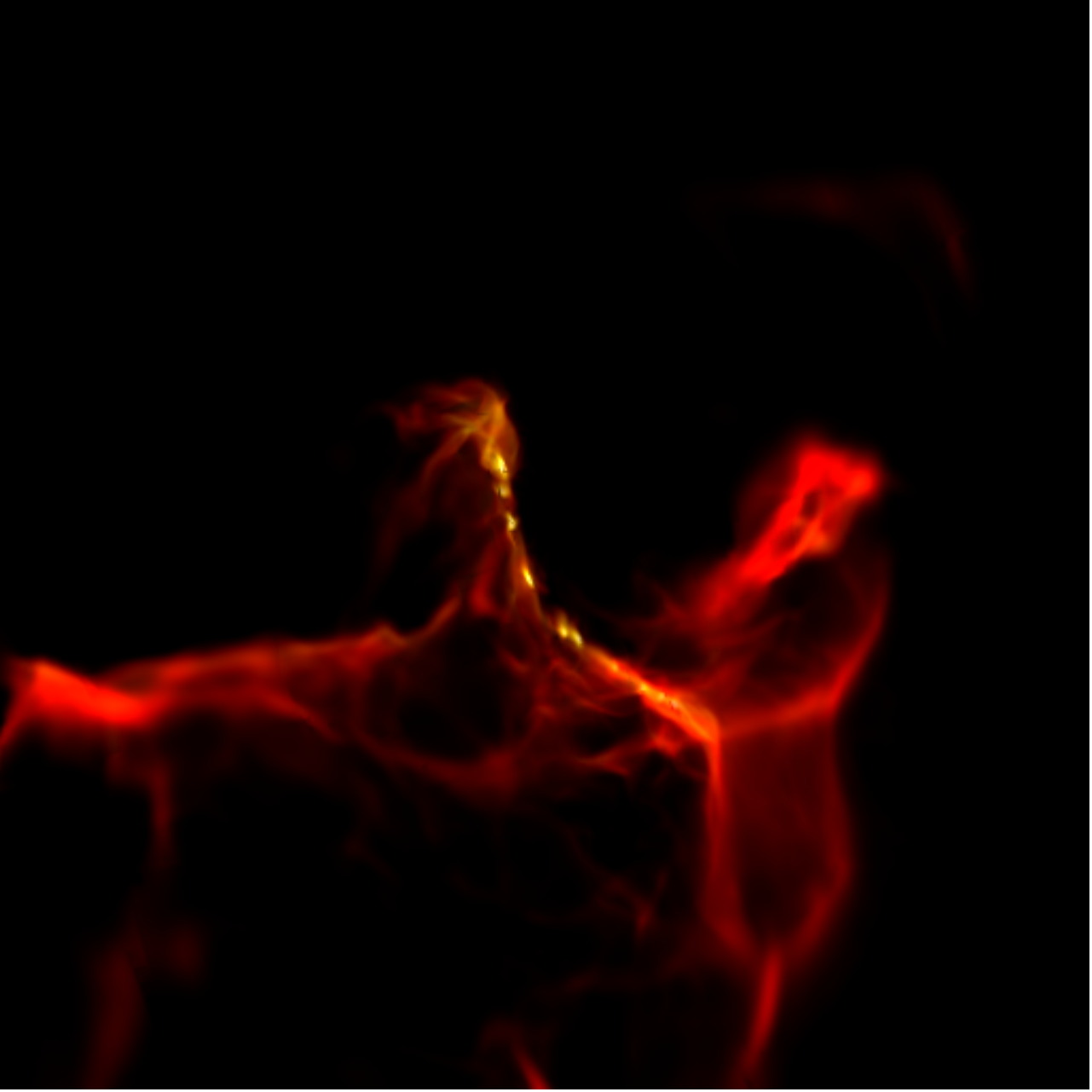}
        \caption{\texttt{M3}, continuous colour table}
        \label{pic_methods_m3e3_0040_3d_v2}
    \end{subfigure}
        \begin{subfigure}{0.32\textwidth}
        \includegraphics[width=\textwidth]{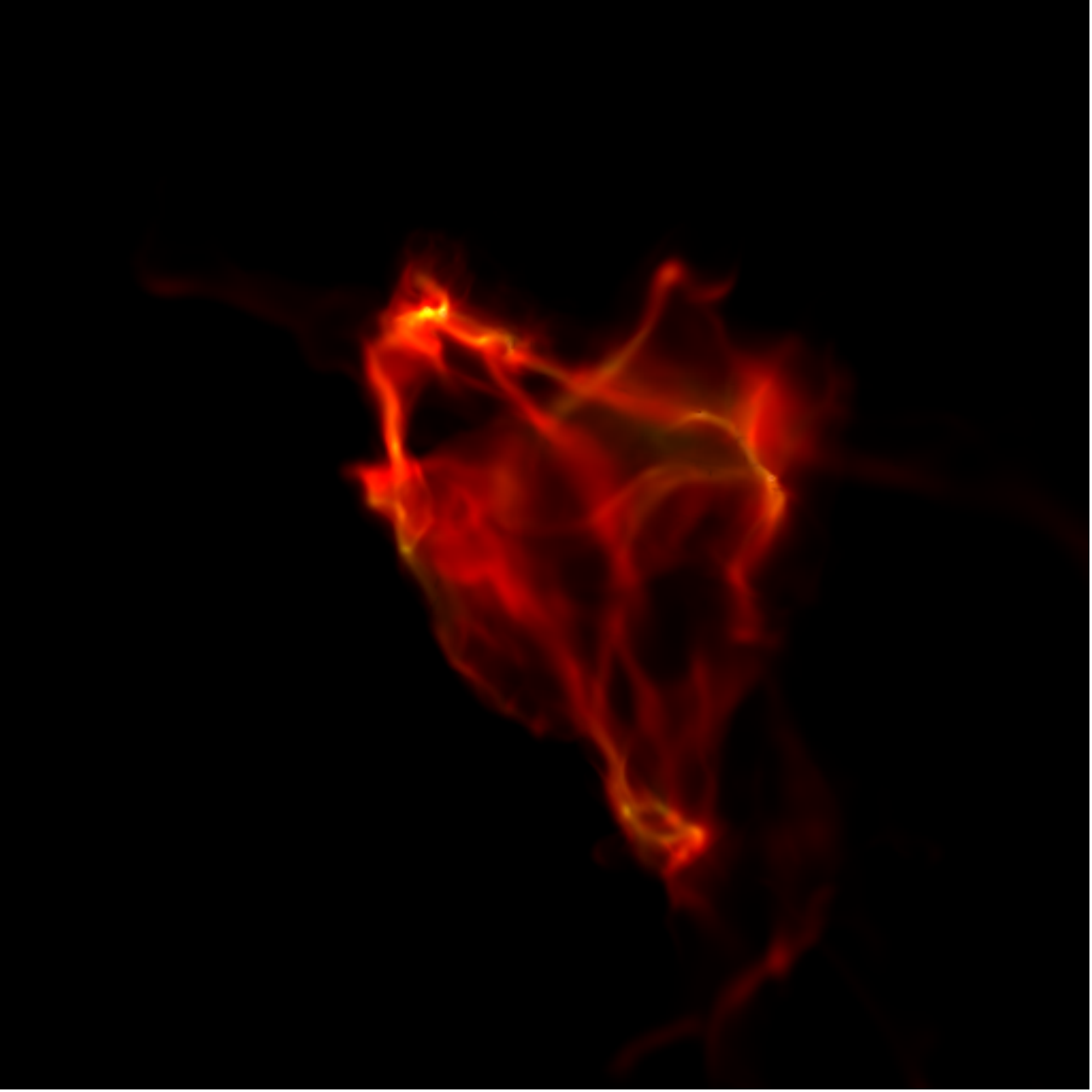}
        \caption{\texttt{M4}, continuous colour table}
        \label{pic_methods_m4e3_0040_3d_v2}
    \end{subfigure}
        \begin{subfigure}{0.32\textwidth}
        \includegraphics[width=\textwidth]{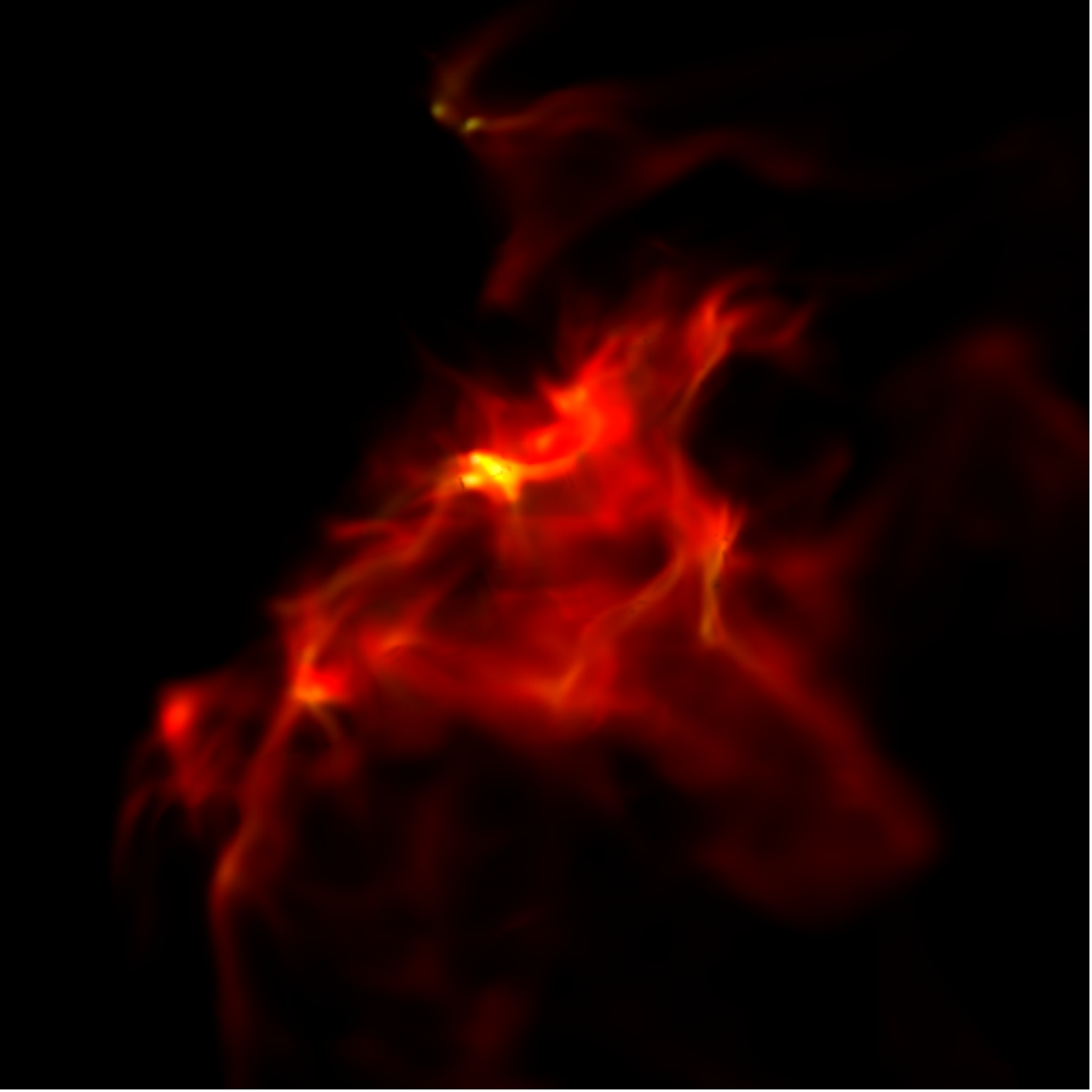}
        \caption{\texttt{M8}, continuous colour table}
        \label{pic_methods_m8e3_0040_3d_v2}
    \end{subfigure}
    
    \caption{Example volume rendered plots showing \texttt{M3} (\textit{left}), \texttt{M4} (\textit{middle}), and \texttt{M8} (\textit{right}) at $t=$4.0~Myr. 
    The colours in the upper panel represent contours at volume densities of 10$^{-23}$~(dark red), 10$^{-22}$ (light red), 10$^{-21}$ (orange), 10$^{-20}$ (yellow), and 10$^{-19}$~g~cm$^{-3}$ (white), while the colour table in the plots in the lower panel is continuous over the same range of volume density.
    }
    \label{pic_methods_volrender_3d}
\end{figure*}

Three dense clouds from the cloud population found in the simulation 
were selected for high-resolution re-simulation by \citet{Ibanez-Mejia2017}. 
Once one of these clouds was identified in the cloud catalogue, a higher-resolution refinement region was defined around the region where the cloud would form in a checkpoint prior to the onset of self-gravity. 
Gas self-gravity was then turned on, and the evolution and collapse of the cloud were followed.
For our investigations, we focus on these three clouds and map a (40~pc)$^{3}$ volume enclosing each cloud with $400\times 400 \times 400$ grid cells, with an effective spatial resolution of $\Delta x_{\rm min}=0.1$~pc.
We consider objects to be fully resolved if their local Jeans length $\lambda_J > 4~\Delta x_{\rm min}$, corresponding to a maximum resolved density at 10~K of $8 \times 10^3$~cm$^{-3}$ \citep[e.g.][Eq.~15]{Ibanez-Mejia2017}.
This means that we can trace fragmentation down to 0.4~pc, but cannot fully resolve objects that form at smaller scales.
The clouds have total masses on the order of $3 \times 10^3$, $4 \times 10^3$, and $8 \times 10^3$~M$_{\odot}$ (hereafter denoted models \texttt{M3}, \texttt{M4}, and \texttt{M8}).
Examples are shown in Fig.~\ref{pic_methods_volrender_3d}.

\subsection{Filament identification}\label{methods_filfinder}

We use a publicly available filament finder that identifies and traces the structures within the model clouds.
We choose \texttt{DisPerSe} \citep{Sousbie2011a} as we can apply it directly to 3D volume density cubes, as well as to 2D column density maps, returning structures based on the same algorithm.
Furthermore, \texttt{DisPerSe} identifies filaments by deriving the gradients within each grid cell and connecting maxima and saddle points with each other, which gives the detected skeletons a mathematical and physical meaning.

We have compared the structures identified by \texttt{DisPerSe} with those detected by other codes (see Appendix \ref{a_filfinder}). 
All of them return similar structures in the densest regions of the clouds, but show pronounced differences in the more diffuse envelopes.
This behaviour has a significant impact on the measured properties of the filaments. 
The consequence is that studies that do not use the same filament finding techniques are not directly comparable unless it has been proven that the identified structures are indeed similar.

In order to obtain results that can be qualitatively compared to observations we follow the approximation of \citet{IbanezMejia2016} and use two thresholds:
(a) a low-density threshold at n$_{\rm th} = $~100~cm$^{-3}$ which defines the volume of the cloud, and corresponds very roughly to the density at which CO emission can first be detected; and
(b) a high-density threshold at n$_{\rm th} = $~5,000~cm$^{-3}$ that points to the denser clumps within the cloud that have a high potential for forming stars \citep[see, e.g.][]{Kainulainen2014}.
We note that we approach our number density resolution (see Sect.~\ref{methods_cloudmodel}) when using this high-density threshold. 
However, we have found that using a lower threshold that is still of the same order of magnitude has no qualitative effect on the structures and properties of the filaments, or their time evolution in general (see also the analysis of the dense gas mass fraction in Sect.~\ref{results_3d_mean} and Fig.~\ref{pic_results_fil3d_dgmf}).

\begin{figure}
        \centering
        \includegraphics[width=0.49\textwidth]{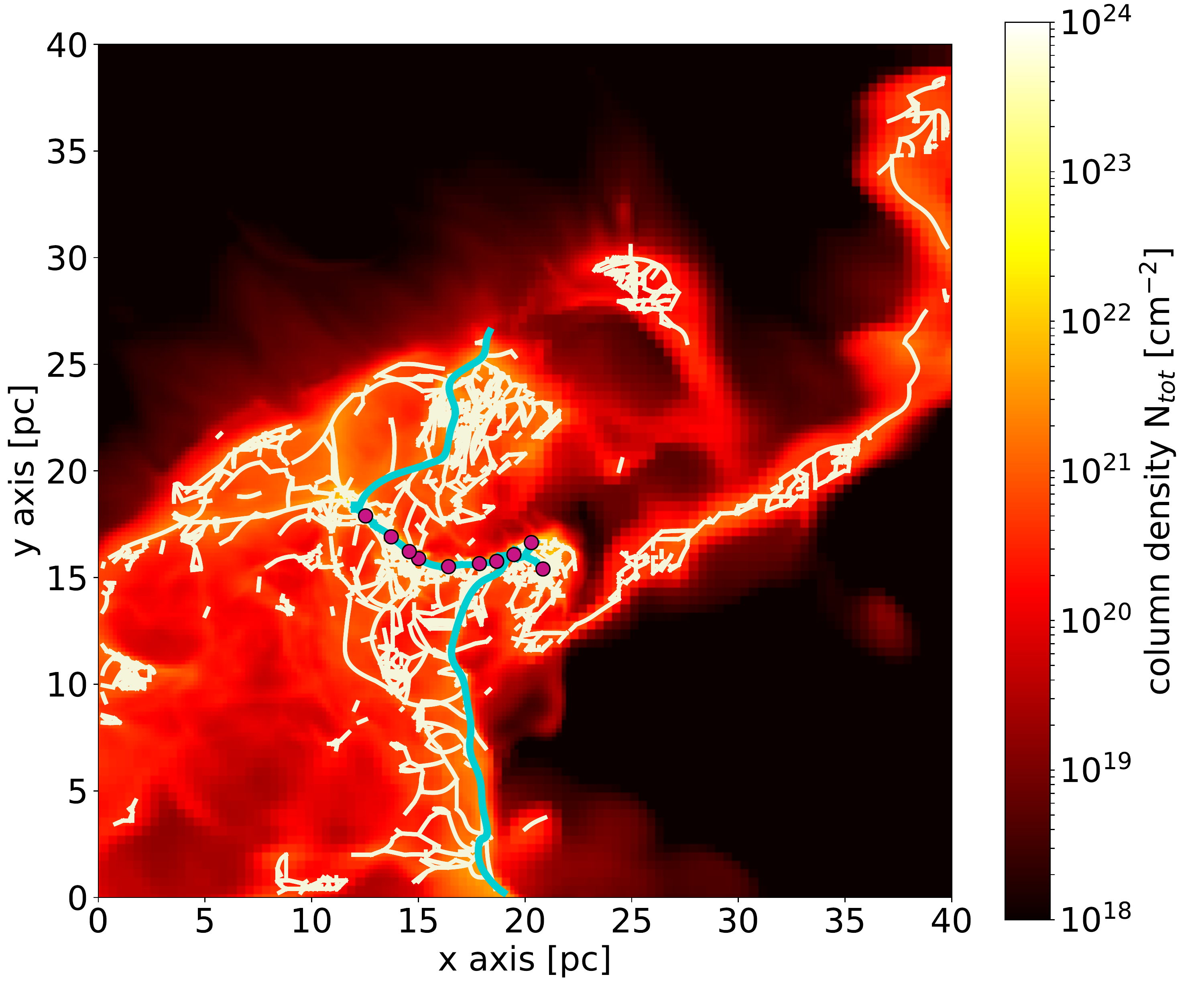}
    \caption{Column density map of \texttt{M3} at $t$ = 3.8 Myr when projected along the z axis with filaments identified using the 3D low density threshold of 100~cm$^{-3}$ (projected, \textit{white lines}), and the 3D high density threshold at 5,000~cm$^{-3}$ (projected, \textit{light blue lines}). The projected positions of identified density fragments are shown with purple dots. }
    \label{pic_results_example_ntot_lowhigh}
\end{figure}

Figure~\ref{pic_results_example_ntot_lowhigh} illustrates with an example how strongly the structures of the identified skeletons depend on the considered threshold.
The map shows a column density map of \texttt{M3} at $t=$3.8~Myr that has been produced by projecting the volume density cube along the z-axis.
The map illustrates a projection of the filament spines identified by DisPerSe in the 3D position-position-position space. 
We note that we only include the spines of the filaments with densities exceeding the density threshold used for their identification, that is, 100 or 5000~cm$^{-3}$. 
This restriction is necessary because of the way \texttt{DisPerSe} identifies filaments. 
\texttt{DisPerSe} only needs a seed point with a (column) density above the specified threshold, and then connects local maxima through tangent lines to the (column) density gradients. 
If no restriction to the (column) density gradients is set, \texttt{DisPerSe} will eventually connect all local maxima, even if they are below the specified (column) density threshold. 
In this work we set no restriction to the (column) density gradient for the filament identification. 
However, once the filamentary network was identified, we kept only those filaments whose spines are above our desired (column) density threshold.

\subsection{Filament properties}\label{methods_mlin}

\texttt{DisPerSe} derives the skeletons of filaments based on the local gradients and user-set thresholds.
It then uses the minimal spanning tree (MST) method to find the nearest neighbour for each element of the skeletons.
Most other filament finders neglect this step and leave it to the user to assemble the individual points to filaments, which may not be as accurate as the algorithm \texttt{DisPerSe} offers. 

In the end, \texttt{DisPerSe} returns the starting points $\mathbf{s_i}$ and end points $\mathbf{e_i}$ of each segment $i$ of an identified skeleton.
With these points, we calculate the length of the filament,
\begin{equation}
        \ell_{\rm tot} = \sum_i | \mathbf{v_i} | = \sum_i | \mathbf{e_i} - \mathbf{s_i} | \, ,
    \label{equ_methods_ltot_def}
\end{equation}
with $\mathbf{v_i}$ being the direction vector pointing from $\mathbf{s_i}$ to $\mathbf{e_i}$.

\begin{figure}
        \includegraphics[width=0.49\textwidth]{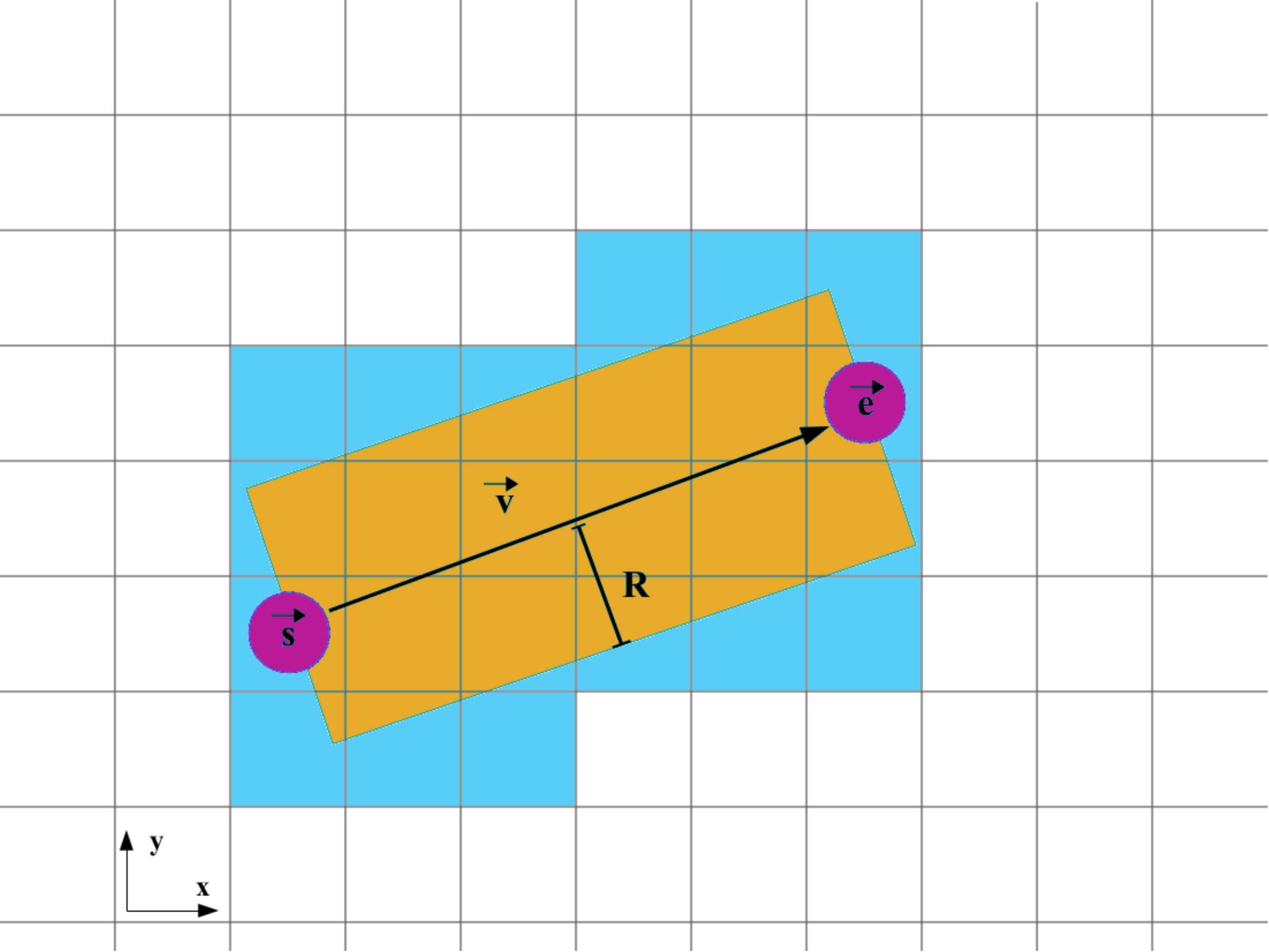}
    \caption{2D sketch describing how filament volume is defined. 
      The purple circles represent the positions of the starting point $\protect\mathbf{s}$, and end point $\protect\mathbf{e}$ on the data grid.
      The filament is defined as the orange cylinder with axis $\protect\mathbf{v}$ and radius $R$.
      The blue area shows the  grid cells actually used to compute the mass, volume, and other quantities.
}
    \label{pic_methods_filident_sketch}
\end{figure}

We define the volume of the filaments as the set of grid cells $\mathbf{x_i}$ that are located along the direction vector $\mathbf{v_i}$ between $\mathbf{s_i}$ and $\mathbf{e_i}$, or have any part that lies within a distance $R$ to these cells.
In Fig.~\ref{pic_methods_filident_sketch} we illustrate this in a 2D example. 
The orange area represents the analytic volume of the filament while the blue area marks the grid cells that are actually used.

The range of filament widths that we find is wide, partly due to the complexity of profiles (e.g. from crossing filaments) that makes it difficult to measure relevant quantities like the full-width at half maximum. 
For simplicity, we assume that the radius of our filaments is $R = 0.3$~pc everywhere.
This value agrees with average filament widths we find when fitting Gaussians to the line profiles of the filaments, although numerical resolution limits filaments in our model from collapsing to smaller extensions.

The enclosed mass of a filament is given by,
\begin{equation}
        M_{\rm tot} = \sum_{\mathbf{x_i} \, \in \, {\rm filament}} \rho(\mathbf{x_i}) \cdot \Delta x_{\rm min}^3 \, ,
    \label{equ_methods_mtot_def}
\end{equation}
where $\rho$ is the mass density within the grid cell $\mathbf{x_i}$; and the line mass by,
\begin{equation}
        M_{\rm lin} = \frac{M_{\rm tot}}{{\ell}_{\rm tot}} .
        \label{equ_methods_mlin}
\end{equation}

\subsection{Fragment identification and properties}\label{methods_frag}

To identify fragments, we use \texttt{astrodendro}\footnote{\url{http://dendrograms.readthedocs.io}}.
This method computes dendrogram trees that represent the hierarchical structure of the underlying matter distribution in order to unravel how sub-structures relate to each other.
This way one can easily identify individual clumps, filaments, and fragments.

We identify fragments as highest-level leaves in the dendrograms derived with a minimal density threshold at \texttt{min\_value}~=~5,000~cm$^{-3}$ and a minimum of \texttt{min\_npix}~=~20~cells.
We define the volume of the fragment as a sphere with a radius of $R_{\rm f}$~=~0.3~pc around the central position of the respective dendrogram leaf.

\section{Results \& Discussion}\label{results}

\subsection{Mean properties and evolution of 3D filaments} \label{results_3d_mean}

We begin our analysis by studying the average properties and time evolution of the filaments we find in our model clouds in 3D, and how those properties relate to the global characteristics of the surrounding clouds.

\begin{figure*}
        \centering
        \includegraphics[width=\textwidth]{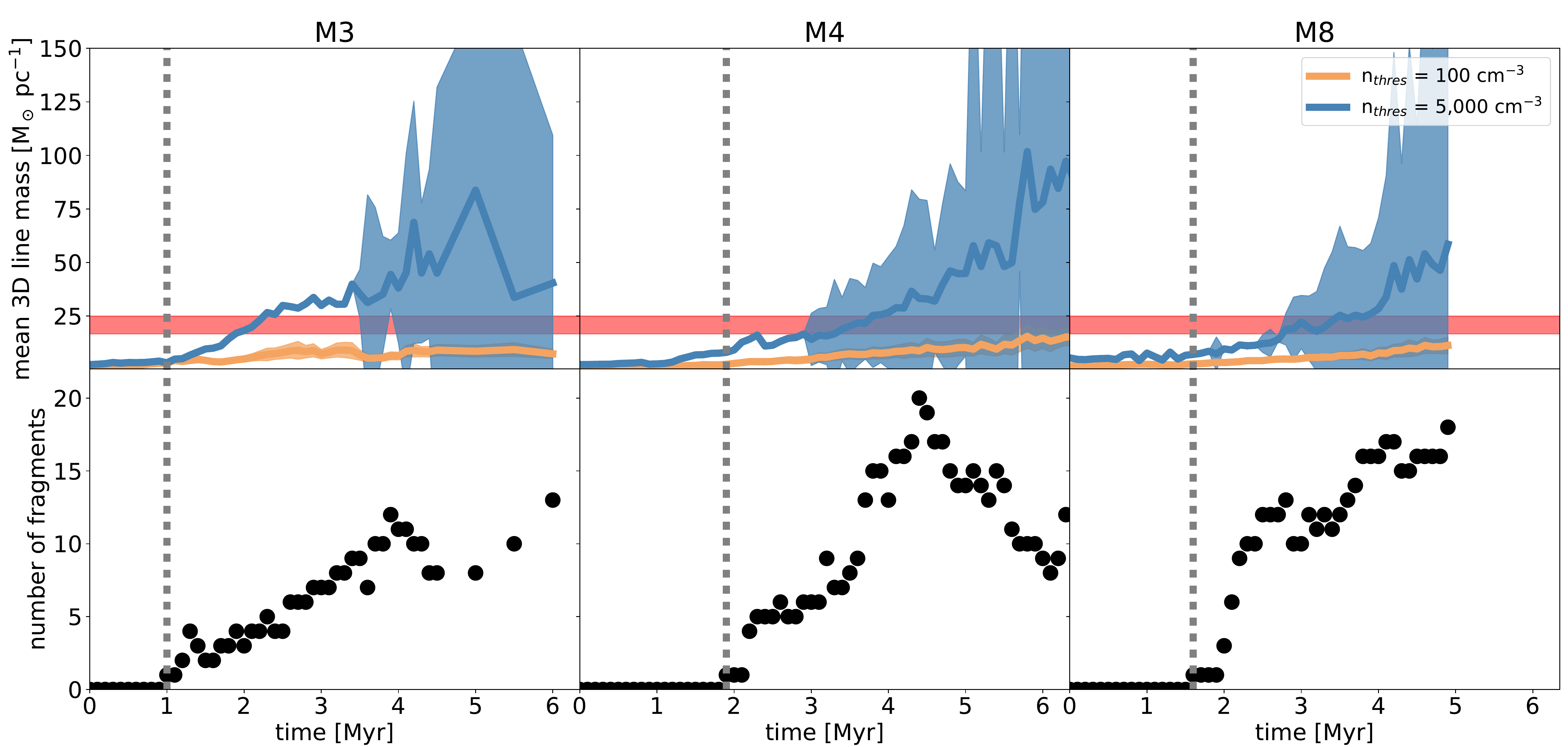}    
        \caption{Average line mass of all filaments $\langle M_{\rm lin,3D} \rangle$ (\textit{top}) and number of fragments (\textit{bottom}) found in models \texttt{M3} (\textit{left}), \texttt{M4} (\textit{middle}), and \texttt{M8} (\textit{right}) as functions of time.
        The coloured areas illustrate the 1$\sigma$ range.
        The red area indicates the critical value of $M_{\rm lin,3D}$ for an isothermal cylinder with temperatures between 10--15~K.
        The grey dashed lines mark the times when the first fragments in each simulation form.
        }
        \label{pic_results_numfrags}
\end{figure*}

In the top panel of Fig.~\ref{pic_results_numfrags} we show the average line mass $\langle M_{\rm lin,3D} \rangle$ of all the filaments identified in each of the three clouds as a function of time $t$ after self-gravity has been activated in the simulations.
We see that the average line masses within all clouds increase with time and with increasing identification threshold $n_{\rm th}$.
This is expected considering the structures we obtain when using different thresholds. 
When we apply a high threshold we obtain compact structures connecting the sites of first fragment formation, where the gas is more concentrated.
In contrast, with a lower threshold we identify a larger number of low-mass structures in more diffuse regions of the clouds (see Fig.~\ref{pic_results_example_ntot_lowhigh}).
As a result, the average line mass of low-density filaments is always lower than when using a higher threshold.

\begin{figure}
        \centering
        \includegraphics[width=0.46\textwidth]{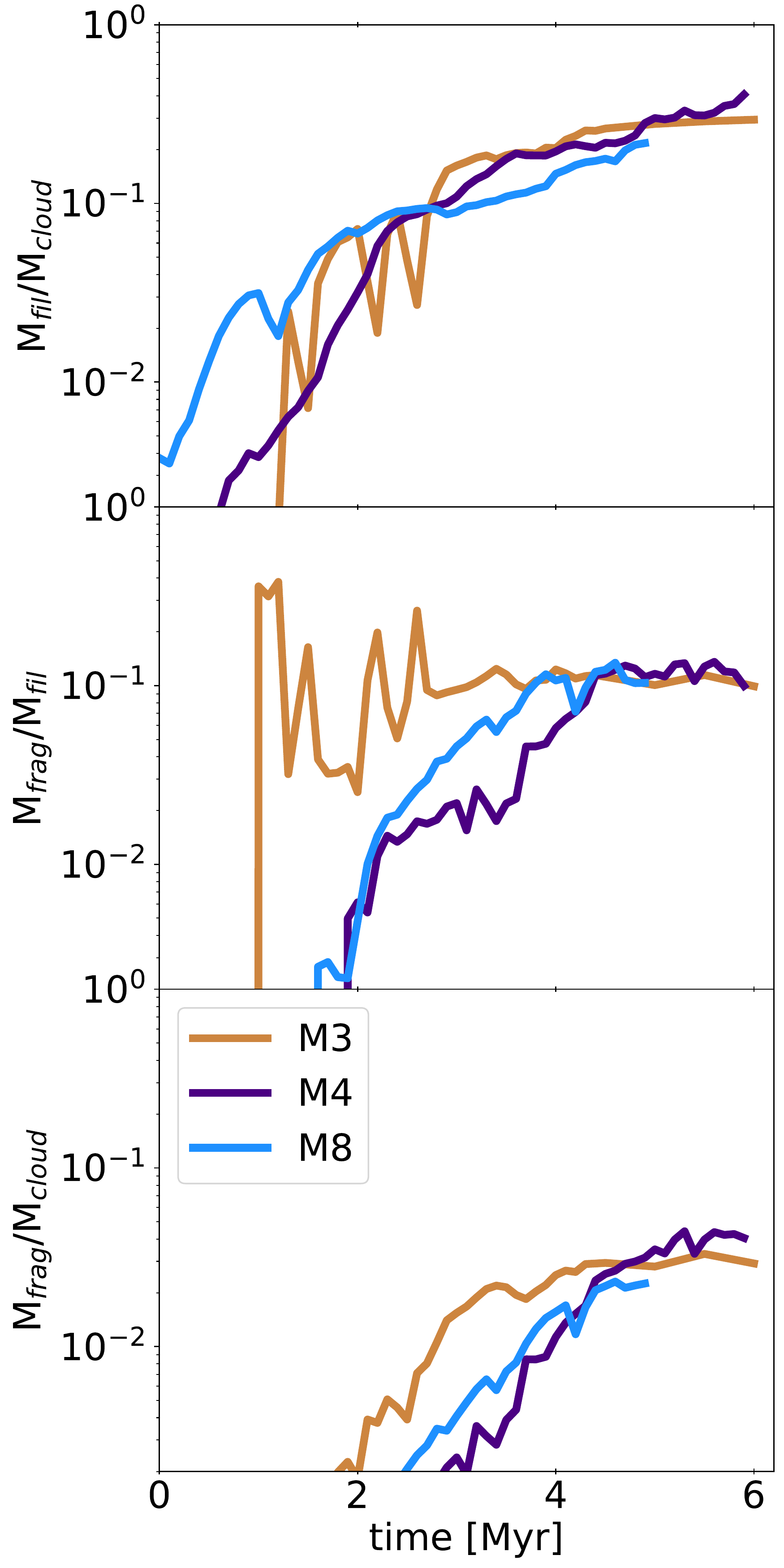}
    \caption{Time evolution of ratios of filament and fragment masses $M_{\rm fil}$ and $M_{\rm frag}$ with respect to cloud masses  $M_{\rm cloud}$ and each other, shown for the objects within \texttt{M3}~(\textit{brown}), \texttt{M4}~(\textit{purple}), and \texttt{M8}~(\textit{blue}).
        The filaments considered have been identified with $n_{\rm th}=100$~cm$^{-3}$. 
    }
    \label{pic_results_massfrac}
\end{figure}

The increase of average line masses over time can be understood from the fact that the clouds formed by compression of gas, due to supernova shocks and global turbulent motions, in the time before our analysis. 
Thus, at $t=0$~Myr, when self-gravity is activated, the clouds are already highly self-gravitating and begin to collapse globally.
As a result, the filaments within the clouds gain more mass with time, which increases their line masses.
Only in the case of M3 does the growth of average line masses of the low-density filaments stagnate after the first megayear. 
However, the average line mass of the high-density filaments  in all clouds steadily increases in time, suggesting that the filaments are collapsing gravitationally.
Figure~\ref{pic_results_massfrac} shows a clearer picture of how the mass within the cloud is distributed, and its time evolution.  
The top panel shows the fraction of mass contained in filaments, $M_{\rm fil}$, compared to the mass of the entire parental cloud, $M_{\rm cloud}$. 
In all clouds, the fraction rapidly and continuously increases in time.
Since the clouds accrete mass from their environments, the growth of the filament mass fraction means that filaments are forming and growing at rates not directly correlated to the evolution of their parental clouds. 
A similar trend is observed in the mass fraction of the fragments with respect to the cloud mass for all three clouds (bottom panel Fig.~\ref{pic_results_massfrac}). 
However, the middle panel showing the mass ratio of filaments to cores shows a steep rise when cores are formed, but then stagnates at later times suggesting that filaments and cores grow at the same rate.

In order to confirm this behaviour, we take a closer look at the dense gas mass fraction (DGMF) of the clouds.
We define the DGMF as the fraction of mass enclosed in cloud cells that contain gas above a given dense gas threshold, n$_{\rm dens}$, compared to the mass of the entire cloud:
\begin{equation}
        {\rm DGMF}(t,n_{\rm dens})  = \frac{ \int_V n(t,\mathbf{x}) \, G(n_{\rm dens}) \,  dV}{\int_V n(t,\mathbf{x}) \, G({\rm 100~cm}^{-3}) \,  dV} \, ,
    \label{equ_results_dgmf}
\end{equation}
with $n$ being the number density at the time $t$ and 
\begin{equation}
        G(n_0) = \begin{cases}
                1 & n(t, \mathbf{x}) \geq n_0 \\
        0 & n(t, \mathbf{x}) < n_0
        \end{cases} .
    \label{equ_results_g_n}
\end{equation}

\begin{figure*}
        \includegraphics[width=\textwidth]{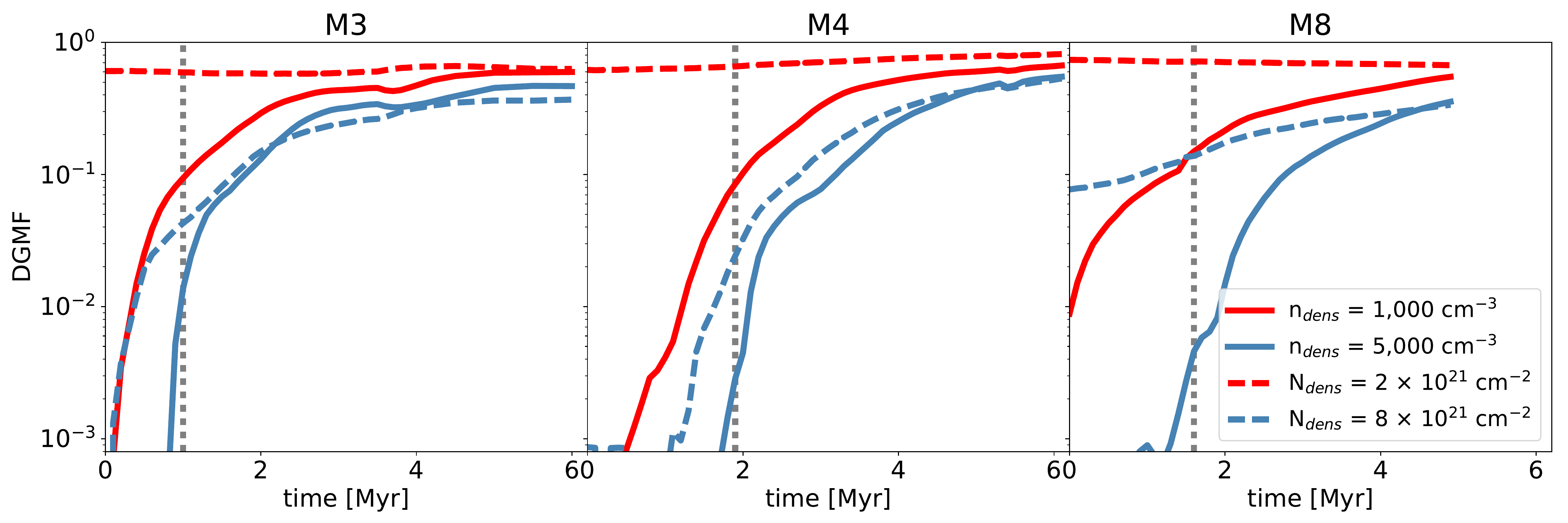}
    \caption{Evolution of DGMF in the model clouds as a function of time. Solid lines give DGMFs measured from 3D volume density cubes, while dashed lines are from 2D column density maps. 
Red lines correspond to dense gas above number and column density thresholds of 1000~cm$^{-3}$ and $2 \times 10^{21}$~cm$^{-2}$, and blue lines correspond to dense gas above 5000~cm$^{-3}$ and $8 \times~10^{21}$~cm$^{-2}$ respectively.
        We see that after self-gravity is activated the gas condenses rapidly, before the growth rate almost stagnates after 2~Myr which corresponds to the clouds' global free-fall times.
                The grey dashed lines mark the times when the first fragments in each simulation form.
    }
    \label{pic_results_fil3d_dgmf}
\end{figure*}

Figure~\ref{pic_results_fil3d_dgmf} shows the evolution of the DGMFs of the three simulated clouds using two different dense gas thresholds, namely $n_{\rm dens}$~=~1,000~cm$^{-3}$ and $n_{\rm dens}$~=~5,000~cm$^{-3}$.

In all clouds we see that the DGMF continues growing until the end of the simulation, with maximal values of 55--70\% for $n_{\rm dens}$~=~1,000~cm$^{-3}$ and 35--55\% for $n_{\rm dens}$~=~5,000~cm$^{-3}$. 
These clouds continue to accrete mass from their surroundings
\citep[see Figs.~3, 8, and 13 in][]{Ibanez-Mejia2017}, so we conclude that these clouds are collapsing faster than they are growing. 

\begin{figure*}
        \includegraphics[width=\textwidth]{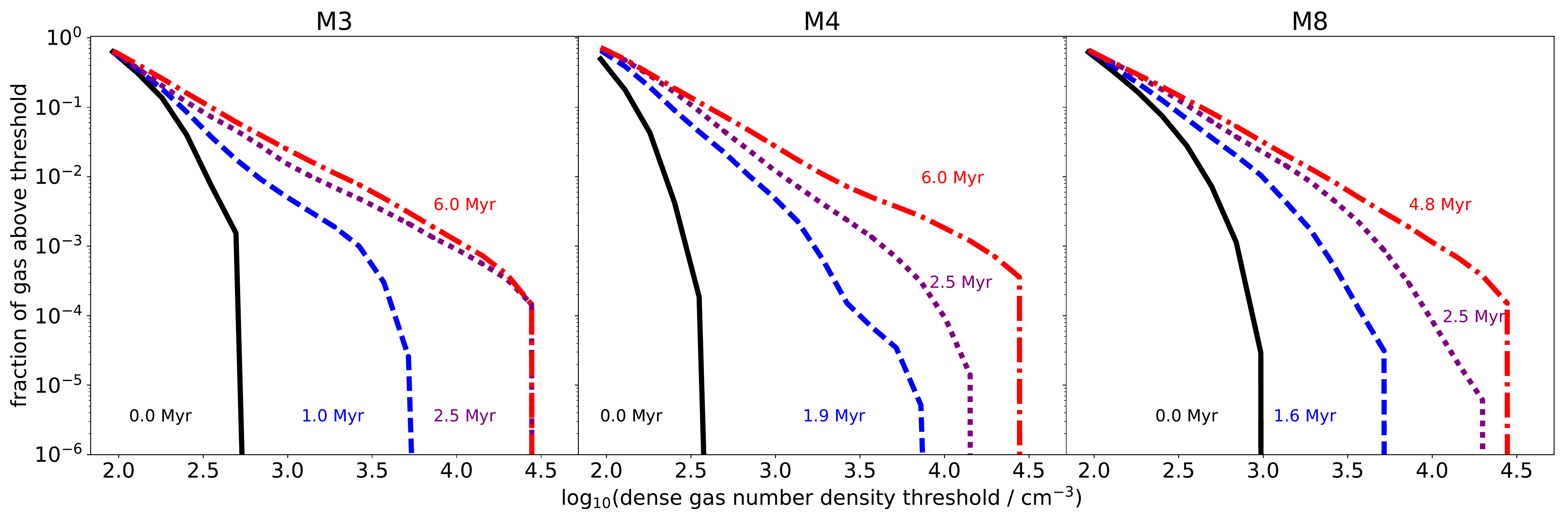}
    \caption{Fraction of gas above the given gas number density threshold (ordinate), $n_{\rm dens}$, within \texttt{M3}~(\textit{left}), \texttt{M4}~(\textit{middle}), and \texttt{M8}~(\textit{right}). 
        The lines show the distribution at the time self-gravity has been activated, $t=$0~Myr (black solid line), the time the first fragments are detected (blue dashed line), $t=$2.5~Myr (violet dotted line), and the final snapshot (red dashed-dotted line).
        }
    \label{pic_results_rho_pdf}
\end{figure*}

We see a similar behaviour in Fig.~\ref{pic_results_rho_pdf}.
In this Figure, we show the fraction of gas above a given number density threshold as a function of this number density threshold, $n_{\rm dens}$, for four individual time steps in the simulations.
In all clouds we see that the maximum gas density grows with time until it reaches the limit of our numerical resolution.
We note that although we reach densities of order 30,000~cm$^{-3}$, we only reliably resolve densities up to 8,000~cm$^{-3}$ .
At the same time, the density distributions become flatter as the clouds evolve.
This means that the clouds collapse in their entirety and compress the enclosed mass into denser substructures such as fragments without necessarily losing their diffuse envelopes.
This likely occurs because of continuing accretion of diffuse gas from the surrounding ISM \citep{Ibanez-Mejia2017}.

In summary, we see that the average evolution of the filaments is influenced by the global kinematics of their parental cloud. In particular, the way that the cloud transforms the mass it accretes from the ISM into dense substructures is related to the formation of fragments within the filaments. 
However, we also see that the properties of the dense gas strongly depend on the parameters used to define and identify it, such as the threshold density. 
This might become important for our key question of how well analytic models evaluate the stability of filaments and predict their fragmentation behaviour.

\subsection{Properties and evolution of individual 3D filaments}\label{results_3d_ind}

In this Section we investigate the evolution of individual filaments and compare the properties of fragmenting filaments with others. 
We confront these properties with the predictions of analytic models.
Table \ref{tab_results_01_prop3d} provides a summary of the properties of the examined filaments.

\begin{table}
        \begin{tabular}{l|cccc}
                Quantity        & min   & mean  & median        & max \\ \hline
                \multicolumn{5}{l}{n$_{\rm th}$ = 100 cm$^{-3}$} \\ \hline
                Number density n [10$^{3}$ cm$^{-3}$]                   &   0.1   &   1.7 &  0.5  & 2079  \\ 
                Gas temperature $T_{\rm gas}$ [K]                       &  10.2   &  11.8 & 11.3  &   16.3        \\ 
                Total length $\ell_{\rm tot}$ [pc]                      &   1.8   & 17.8  & 11.2  & 133.2         \\ 
                Total mass $M_{\rm tot}$ [M$_\odot$]                    &   0.4   & 107.0 &  8.2  & 1138          \\ 
                Line mass $M_{\rm lin,3D}$ [M$_\odot$ pc$^{-1}$]        &   0.2   &   2.3 &  0.8  &   17.3        \\ \hline 
                \multicolumn{5}{l}{n$_{\rm th}$ = 5,000 cm$^{-3}$} \\ \hline
                Number density n [10$^{3}$ cm$^{-3}$]                   &  2.5    &  26.0 &  13.5 & 2116          \\ 
                Gas temperature $T_{\rm gas}$ [K]                       & 10.2    &  11.8 &  11.3 &   16.3        \\ 
                Total length $\ell_{\rm tot}$ [pc]                      &  1.8    &   5.6 &   5.0 &   15.0        \\ 
                Total mass $M_{\rm tot}$ [M$_\odot$]                    & 28.8    & 553.6 & 409.6 & 2553          \\ 
                Line mass $M_{\rm lin,3D}$ [M$_\odot$ pc$^{-1}$]                & 13.2    &  97.0 &  73.3 &  707.6        \\ \hline 
        \end{tabular}
        \caption{Summary of properties of the 3D filaments.}
        \label{tab_results_01_prop3d}
\end{table}

\begin{figure*}
        \centering
        \begin{subfigure}{0.98\textwidth}
        \includegraphics[width=\textwidth]{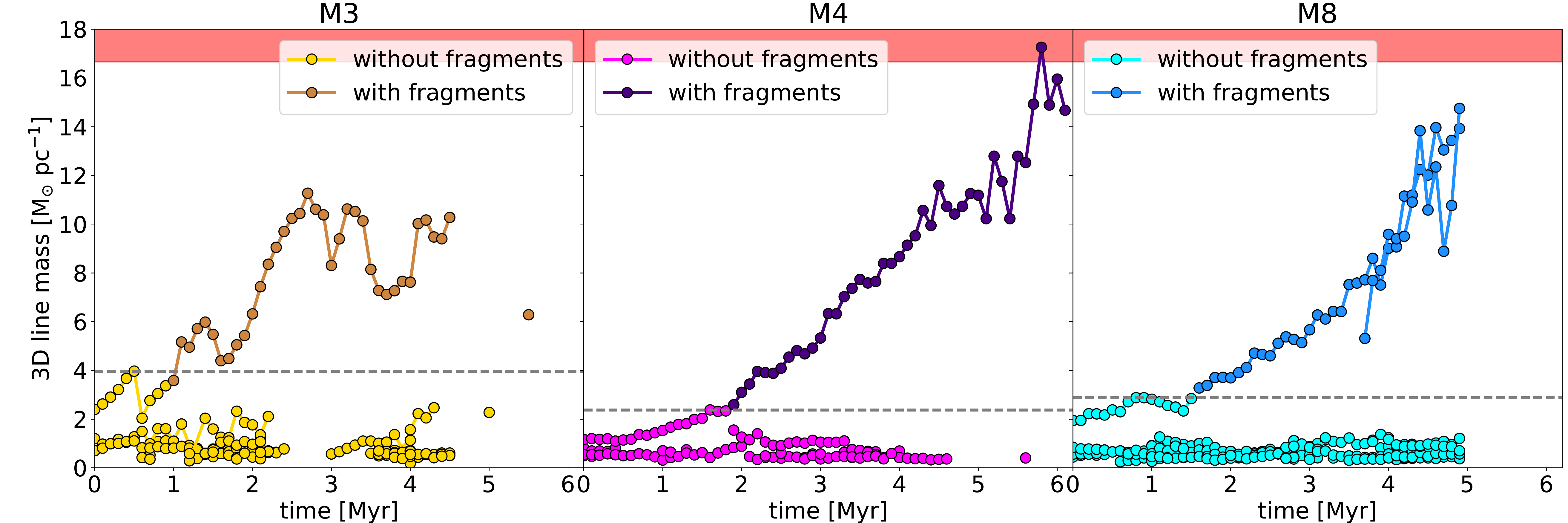}
        \caption{Filaments identified with threshold of 100 cm$^{-3}$.}
        \label{pic_results_filsep_low}
    \end{subfigure}
    
        \begin{subfigure}{0.98\textwidth}
        \includegraphics[width=\textwidth]{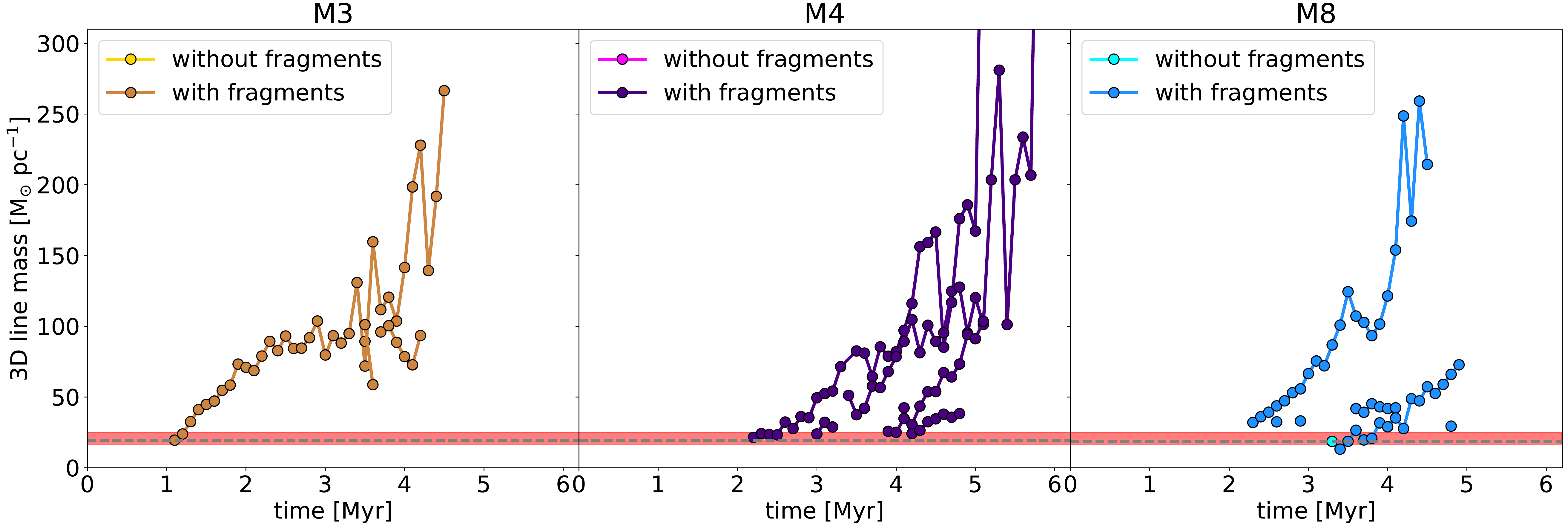}
        \caption{Filaments identified with threshold of 5,000 cm$^{-3}$.}
        \label{pic_results_filsep_high}
    \end{subfigure}
    
    \caption{Evolution of line mass of individual filaments.
        Each line represents a single filament, with dots showing its line mass at each output time step.
        The light dots illustrate the time steps when the respective filament does not contain any fragments, the darker dots show these time steps  when the filament contains at least one fragment.
        The red area illustrates the critical line mass at which isothermal filaments with temperatures between 10 and 15~K become unstable against collapse according to the \citet{Ostriker1964b} model.
        The grey dashed line shows the maximum line mass of filaments {without} fragments in the simulation.
    }
    \label{pic_results_filsep}
\end{figure*}

In Fig.~\ref{pic_results_filsep} we show the line masses of {individual} filaments as a function of time.
In the case of the low-density filaments (Fig.~\ref{pic_results_filsep_low}), the transition from filaments without embedded fragments to those with fragments occurs at very low line masses (4.0, 2.4, and 2.9~M$_\odot$~pc$^{-1}$ for \texttt{M3}, \texttt{M4}, and \texttt{M8}, respectively).

We compare the line masses at which fragmentation occurs in the model with the criterion for cylindrical filaments being in hydrostatic equilibrium.
In particular, we focus on the model by \citet{Ostriker1964b} which describes a filament as an infinitely long, isolated, isothermal cylinder filled with self-gravitating gas that is in balance between gravity and thermal pressure. 
In this model, the equilibrium configuration is uniquely characterised by the critical line mass
\begin{equation}
        M_{\rm lin,crit} = \frac{2 \, c_s^2}{G} \approx 16.6 \, \left( \frac{{\rm T_{\rm gas}}}{{\rm 10 \, K}} \right) \,  {\rm M}_{\odot} \, {\rm pc}^{-1} \, ,
        \label{equ_methods_mlinecrit}
\end{equation}
with $c_s$ being the sound speed of the gas, $G$ the gravitational constant, and $T_{\rm gas}$ the gas temperature.

According to \citeauthor{Ostriker1964b}, a cylindrical filament is only thermally supported against collapse if its line mass remains below this critical value.
Otherwise, initially small perturbations within the filament can grow under the influence of self-gravity, which then leads to radial collapse and fragmentation, as \citet{Nagasawa1987}, \citet{Inutsuka1992}, \citet{Fiege2000b}, and \citet{Fischera2012a} describe in their fragmenting cylinder models.

Thus, the cylindrical fragmentation model has become a commonly used description of a filament and set of initial conditions of its fragmentation.
The key question is whether or not the evolution of the filaments in our simulations follows this criterion.
Can the equilibrium configuration be regarded as a realistic initial condition for fragmentation?
Is the equilibrium configuration, in fact, ever reached?

To address these questions, we mark the range of the analytic values for the critical line masses for typical gas temperatures between 10 and 15~K with red areas in Fig.~\ref{pic_results_filsep}.
We see that the low-density filaments in our samples start fragmenting at line masses far below the predicted critical values, and hardly reach such high line masses even at later stages in their evolution.

This suggests that the approximations of the cylindrical fragmentation model fail here.
We see many differences between our filaments and those in the analytic model.
In particular, our filaments are neither isolated, nor in hydrostatic equilibrium, nor cylindrically symmetric, as studies by, for example, \citet{Gomez2014} emphasise.
Rather, they are part of a hierarchically collapsing cloud, and interact with each other; for example~by crossing each other or accreting gas.
Thus, the filaments are both subject to external pressure and may have large density perturbations that are outside the regime of linear growth.

Furthermore, studies \citep[e.g.][]{Nagasawa1987,Inutsuka1992,Fiege2000b,Fischera2012a} have found that even filaments that are subcritical in terms of line mass can fragment.
They show that fragmentation itself does not show any clear imprints of the forces that originally formed the fragment (e.g. shock waves, or cloud-cloud collisions).
The conclusion is that there is no prediction for a threshold below which the fragmentation of filaments is prevented.

The situation changes when we look at the high-density filaments (Fig.~\ref{pic_results_filsep_high}).
Here we observe maximal line masses of filaments without fragments between 18 and 20~M$_\odot$~pc$^{-1}$ in all clouds.
These transitional line masses appear to be in agreement with the critical line mass of the cylindrical fragmentation model.
The reason for this is that the properties of the filaments here are more likely comparable to those of the cylindrical fragmentation model since they are more likely isothermal, straighter, and closer to the sites where overdensities form compared to more diffuse filaments. 
Consequently, the objects we study here agree better with the properties of the analytic cylinders.
However, we detect only one dense filament without embedded fragment for one single time step (in \texttt{M8}) and, thus, lack a statistically meaningful sample to draw final conclusions about the capability to predict the fragmentation behaviour of high-density filaments.
From the analysis of the low-density filaments, though, we see that the configuration represented by the analytic model of cylindrical fragmentation is not universally part of the evolution of filaments in the simulations.

We conclude that a simple cylindrical model as described by \citet{Ostriker1964b} and the cylindrical fragmentation model does not represent the typical initial conditions for the fragmentation of a filament in our molecular cloud simulations.
Therefore, it is not a complete model for evaluating the stability of filaments.
To predict the fragmentation of a filament, a more complex model is essential; one that not only considers the balance between internal self-gravity and thermal pressure, but also other forces, such as external hydrostatic, turbulent, or magnetic pressure. 
Furthermore, connecting such a model with observations needs to take into account that the properties of the filaments derived strongly depend on the parameters used to identify the filaments (as well as the filament-finder code used, as discussed in Appendix~\ref{a_filfinder}).
This also means that whether or not observed filaments fragment as predicted by a cylindrical fragmentation model is strongly influenced by the identification parameters, as one can always choose parameters in a way that it perfectly suits the model. 
In our case, this is represented by the high-density filaments that only then contain fragments after exceeding the critical line mass.
Since there is no unique, universal, and physically motivated definition of what filaments are, the fragmentation models are not universal themselves.

\subsection{Properties and evolution of fragments}\label{results_frag}

In this Section, we discuss the properties of the fragments we have detected within our clouds using \texttt{astrodendro} (Sect.~\ref{methods_frag}), including their time evolution, and connection to their parental filaments and to each other.
The main properties of the fragments are summarised in Table~\ref{tab_results_02_propfrag} showing that those fragments have similar properties as dense regions and condensations in star-forming regions \citep{Bergin2007}.

\begin{table}
        \begin{tabular}{l|cccc}
                Quantity & min & mean & median & max \\ \hline
                Number density n [10$^{3}$ cm$^{-3}$]   &  4.1  & 27.1  & 27.1    & 2136          \\ 
                Total mass $M_{\rm tot}$ [M$_\odot$]    &  2.1  & 48.6  & 48.6    &  744          \\ 
                Gas temperature $T_{\rm gas}$ [K]       & 10.0  & 10.1  & 10.1    &   48.6        \\ 
                Closest neighbour separation [pc]           &  0.4  &  1.2  &  1.0    &    4.0        \\
                Virial parameter at first detection     &  0.5  &  3.5  &  3.5    &    7.7        \\  \hline
        \end{tabular}
    \caption{Summary of properties of fragments.}
    \label{tab_results_02_propfrag}
\end{table}

In order to find evidence for  the fragments in our simulation forming by gravitational fragmentation following the cylindrical fragmentation criterion, 
we make a first estimate of how gravitationally bound our fragments are.
For this, we compute their virial parameters, given by \citep{Bertoldi1992}:
\begin{equation}
        \alpha = \frac{{\rm 2} E_{\rm kin}}{E_{\rm pot}} = \frac{\mathrm{5} \sigma R_{\rm f}}{G M_{\rm f}}
    \label{equ_results_virial_parameter}
,\end{equation}
\noindent for each of the fragments at any time step.
Here, $G$ is the gravitational constant, $M_{\rm f}$ the enclosed mass of the fragment, $R_{\rm f}$~=~0.3~pc its radius and $\sigma$ its velocity dispersion, which we calculate following Eqs.~(9)~\&~(10) in \citet{IbanezMejia2016}:
\begin{equation}
        \sigma^{\rm 2} = \frac{1}{3} \, \frac{\sum_{|\mathbf{x}-\mathbf{x_0}| < {\rm R}_f} \, \rho(\mathbf{x}) \, (\mathbf{v} - \bar{\mathbf{v}})^2}{\sum \rho(\mathbf{x})} + \bar{c}_{\rm s}^2 ,
\end{equation}
with $\mathbf{x_0}$ being the central position of the fragment, $\bar{\mathbf{v}}$ the average gas velocity vector, and $\bar{c}_{\rm s}$ the average sound speed.
We note that this estimate only takes the ratio between volumetric kinetic and gravitational energy into account.
Other terms, such as magnetic fields or surface thermal or kinetic pressure \citep{McKee1992}, are neglected. 
This is a common approximation used in both observational and theoretical studies, although \citet{Ballesteros-Paredes2006} suggest that these additional terms can be as important for objects embedded within molecular clouds as the balance of volumetric kinetic and gravitational energy. 

\begin{figure}
        \includegraphics[width=0.49\textwidth]{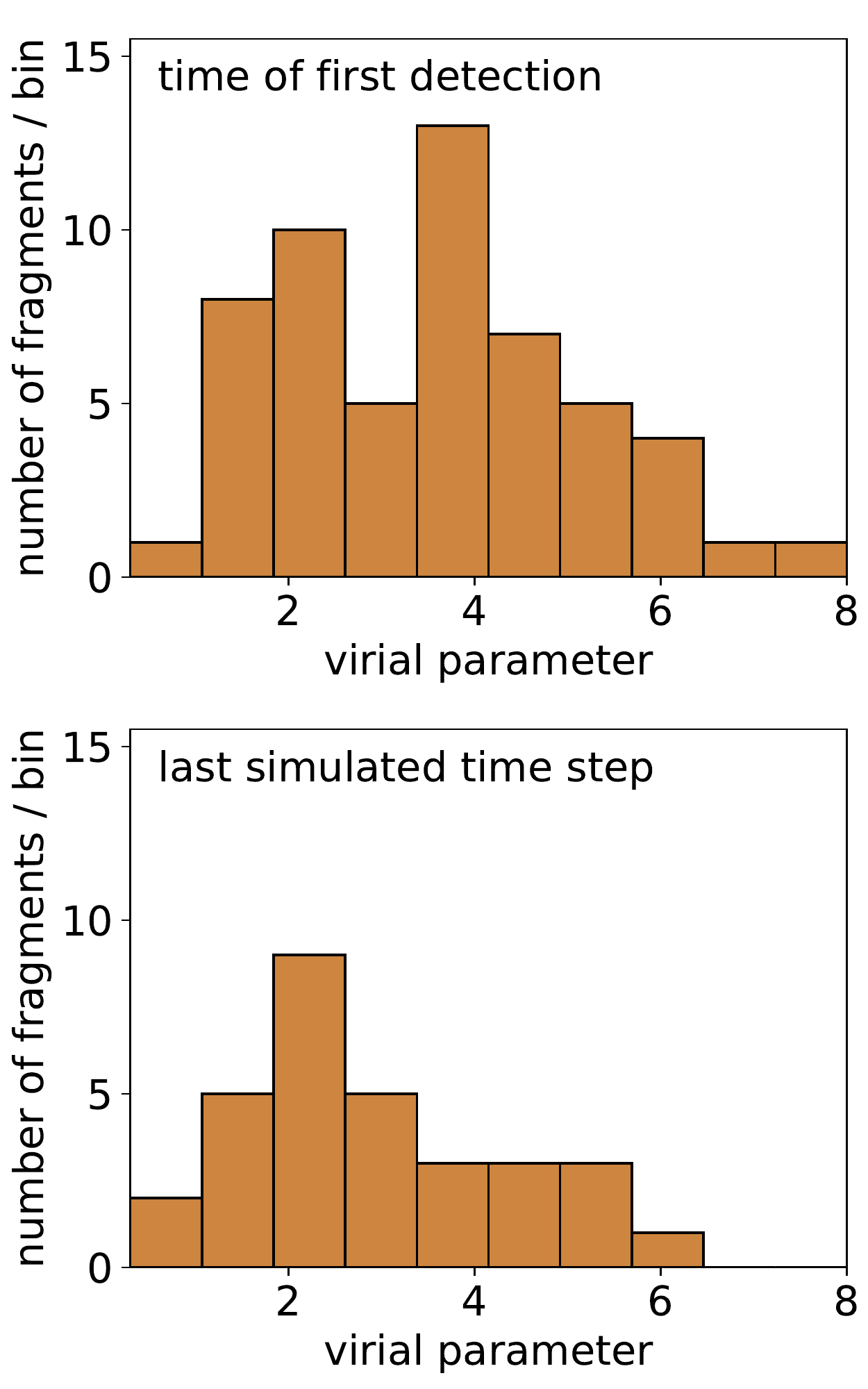}
    \caption{Histogram of volumetric virial parameters of fragments (from all three clouds) at the time when they were initially detected (\textit{top}) and in the last simulated time step (\textit{bottom}).
    }
    \label{pic_results_dendro_alpha}
\end{figure}

In the top row of Fig.~\ref{pic_results_dendro_alpha} we show a histogram of virial parameters measured for the fragments at the time when they have been first detected.
According to Eq.~(\ref{equ_results_virial_parameter}), the fragments are gravitationally bound when $\alpha < 2$. 
This criterion is fulfilled by only 20\% of the fragments.
The majority of fragments have virial parameters close to the mode of 3.5.
We argue that these fragments may nevertheless be bound objects because of the terms not accounted for by Eq.~(\ref{equ_results_virial_parameter}), particularly the surface terms resulting from the collapse of the surrounding cloud.
Furthermore, we see that the virial parameters decrease as the fragments evolve, which means that they become more bound.
The histogram in the bottom row of Fig.~\ref{pic_results_dendro_alpha} confirms this as it shows that most of the fragments have virial parameters around 2 in the last simulated time steps.

We note that, as explained by \citet{IbanezMejia2016,Ibanez-Mejia2017}, we can only give lower limits on the velocity dispersions, and thus the virial parameters, since we under-resolve the turbulence on small scales in the simulations.
Resolving the subgrid scale turbulence may increase the energy by 25\% \citep{Ibanez-Mejia2017}, but likely won't prevent the fragments from beginning to collapse before the criteria for cylindrical fragmentation are fulfilled.
This consequently means that the fragmentation must be driven by neglected terms in the virial equation that can produce significant differences in the estimation of the fragments' boundness.

In the bottom row of Fig.~\ref{pic_results_numfrags} we plot the number of identified fragments as a function of time.
We see that the number of fragments overall increases with time for each simulation.
However, there are cases when the number of fragments drops. 
The missing fragments are either disrupted, for example by shock waves or intracloud turbulence, or merge with each other, meaning that they approach each other too closely ($<$0.4~pc) to be distinguishable with our resolution. 

We see that the first fragments form within the first 2~Myr, corresponding to 25--50\% of the clouds' free-fall times and the period when the growth of the DGMF is steepest (see Fig.~\ref{pic_results_fil3d_dgmf}).
This indicates that the formation of fragments is primarily dominated by the compression of dense gas within the cloud.
This is verified by the ratio of mass contained within the fragments, $M_{\rm frag}$, relative to the mass of the entire cloud, $M_{\rm cloud}$, shown in the bottom panel of Fig.~\ref{pic_results_massfrac}.
The $M_{\rm frag}/M_{\rm cloud}$ ratios evolve in a similar fashion as the DGMF.
Compared to the more constant growth of $M_{\rm cloud}$ or $M_{\rm fil}$, the ratio $M_{\rm frag}/M_{\rm cloud}$ increases rapidly before stagnating around 0.03.
Interestingly, if we approximate the star formation efficiency per free-fall time of our clouds with this ratio, the value agrees with typically observed efficiencies in molecular clouds \citep[e.g.][]{Krumholz2007}.
This is consistent with the determination of star formation efficiency by the dynamics of gravitational collapse, though our relatively small sample and lack of feedback modelling does not allow definitive conclusions to be drawn.

Subsequently, we follow the evolution of the mass contained in the fragments relative to the mass contained in the filaments, $M_{\rm frag}/M_{\rm fil}$.
We plot the ratios in the middle panel in Fig.~\ref{pic_results_massfrac}. 
In all clouds, we see that the $M_{\rm frag}/M_{\rm fil}$ ratio increases rapidly within the first $\sim$2.5~Myr after the first fragments have formed, reaching maximal values of 15--40\%. 
This demonstrates that the fragments accrete mass from their parental filaments efficiently as long as there is a sufficient gas reservoir accessible, as is the case at the beginning of the simulations.
In doing so, they take up a significant fraction of the filaments' masses.

Another question that has been recently discussed in the literature is whether prestellar cores form in a regular pattern within filaments.
Observations suggest that cores condense at regular intervals along their parental filaments \citep[e.g.][]{Jackson2010,Hacar2011}.
The mean separations between the cores appear to correlate with the properties of the respective filament, ranging from a few tenths to several parsecs.
These observations seem consistent with theoretical models of periodic fragmentation \citep{Ostriker1964a,Nagasawa1987,Inutsuka1992,Fischera2012a}.
The instabilities causing fragmentation in these models have unique modes that depend on the initial conditions of the filament.
The wavelengths of these modes then define the mean separations between the forming cores.
However, other studies, both observational \citep[e.g.][]{Enoch2006,Gutermuth2009} and theoretical \citep[e.g.][]{Seifried2015,Clarke2017}, demonstrate that periodic fragmentation only occurs under special conditions (such as supersonic, purely compressive turbulent motions), if at all. 
In reality the conditions within the filaments are not as uniform as assumed by the models, so observed core patterns may also be the result of overlapping fragmentation modes.
According to those studies, filaments commonly fragment in a disordered, cluster-like fashion. 

\begin{figure*}
        \includegraphics[width=\textwidth]{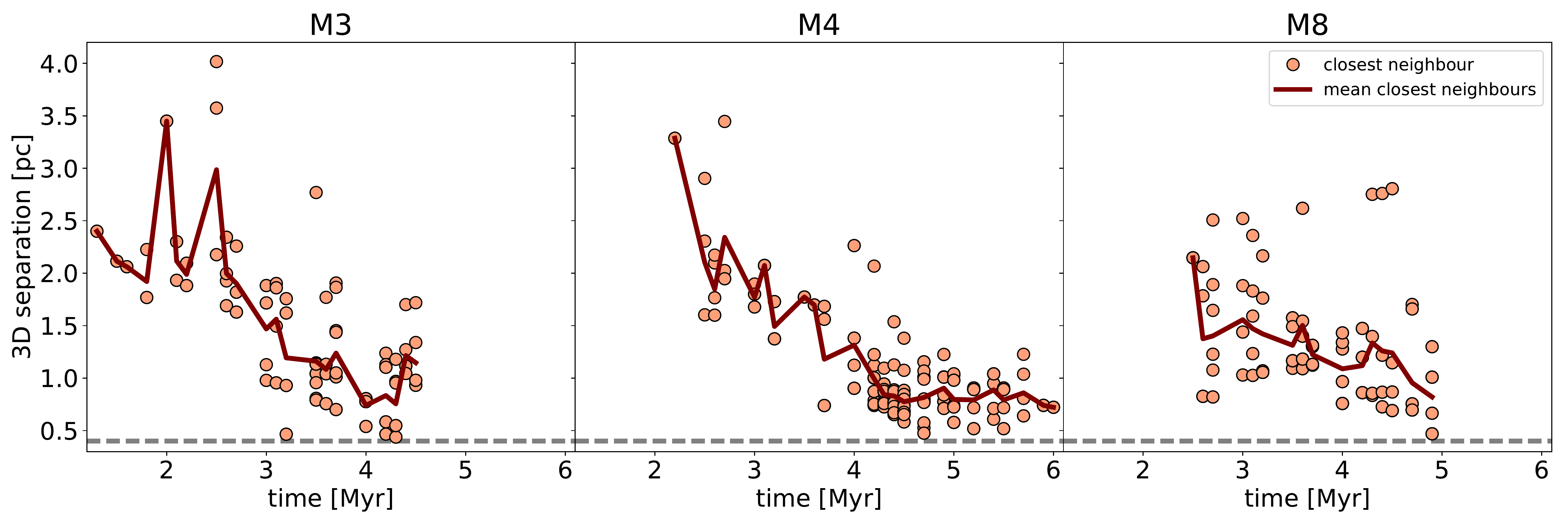}
    \caption{3D separations between fragments and their individual closest neighbours (orange dots) within the same filament within \texttt{M3} (\textit{left}), \texttt{M4} (\textit{middle}), and \texttt{M8} (\textit{right}) at each time step.
        The line illustrates the mean separation between closest neighbours over time.
        The grey dashed line represents our resolution limit at 0.4~pc below which we cannot distinguish individual cores from each other.
    }
    \label{pic_results_fragments_meansep}
\end{figure*}

Using our data, we test whether or not the fragments in our sample form with uniform separations.
We note that we can only detect separations that are larger than 0.4~pc due to the 0.1~pc resolution of the data grid and the 0.3~pc radius we assume for the fragments.
We plot the separations in 3D space of the individual fragments to their individual closest neighbour within the same filament as a function of time in Fig.~\ref{pic_results_fragments_meansep}.
We see that the fragments form, on average, at distances exceeding 2~pc from their closest neighbour at the beginning of the fragmentation process, but approach each other with time down to $<$1~pc or even merge (below 0.4~pc).

\begin{figure*}
        \includegraphics[width=\textwidth]{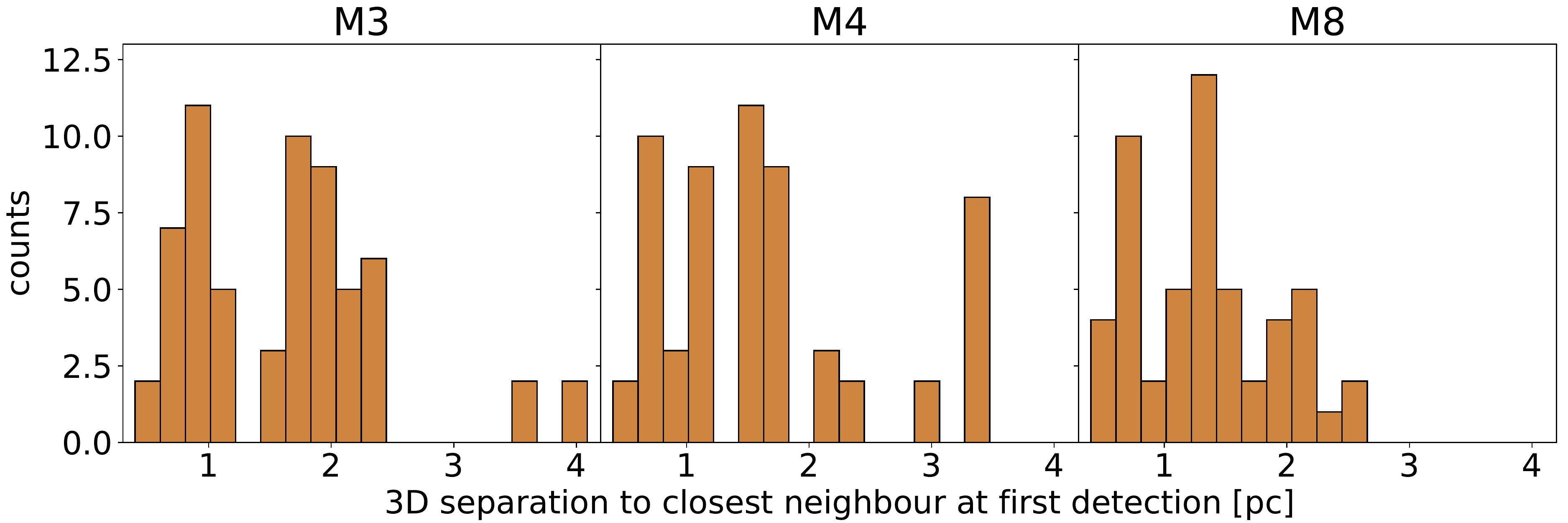}
    \caption{Histogram of separations of fragments within \texttt{M3} (\textit{left}), \texttt{M4} (\textit{middle}), and \texttt{M8} (\textit{right}) to their closest neighbour at the time when the fragment was detected for the first time.
    }
    \label{pic_results_fragments_hist_sep_first}
\end{figure*}

To answer the question of whether there is a typical fragmentation scale we need to consider the separations between closest neighbours at the moment the fragments form.
These are summarised in Fig.~\ref{pic_results_fragments_hist_sep_first}. 
If there were a typical separation we should see a significant peak at that particular scale length, or a sequence of aliased peaks with equal separations. 
Looking at the histograms, we might argue that we see such sequences in \texttt{M3} and \texttt{M8} with typical separations at 0.9 and 0.6~pc, respectively.
Both numbers exceed the local Jeans length by a substantial factor ($\sim$0.1~pc), but are only a factor of 2.3--1.5 larger than our resolution limit. Furthermore, in the case of \texttt{M4}, we do not see any significant peak in the distribution.
However, the number of fragments in the samples is too low to make a solid statement about the existence of  a universal fragmentation scale length, and if it depends on the local physical conditions only.

\subsection{Properties and evolution of filaments in 2D}\label{results_2d}

In the previous Subsections, we studied the properties and evolution of the filaments and fragments that we identified based on the full 3D simulation data.
In observations, however, such 3D data are not available \citep[see, however,][ for a method to reconstruct the volume density distribution]{Kainulainen2014}.
Instead, we observe the filaments projected onto the 2D plane of the sky.
This raises the questions of what we would observe if we apply our methods to the projected data and how these results compare to the results from 3D data.

In this Section, we approach these questions by projecting our 3D volume density cubes onto 2D column density maps.
We project along the three major axes $x$,~$y$, and~$z$, in order to account for line-of-sight (LoS) specific variance. 
We note that these maps lack any additional observational effects, such as noise, beam, or optical depth effects, and are thus only idealised approximations to real observations.
However, such maps give the best case scenario to test the 3D-2D correspondence.
Since we neglect opacity, our column density maps are primarily comparable to optically thin (sub-)mm dust observations. 
To produce comparable synthetic spectral line emission maps, we would need to consider chemical abundances, as well as gas velocity, which is beyond the scope of this paper.

\begin{figure}
        \includegraphics[width=0.49\textwidth]{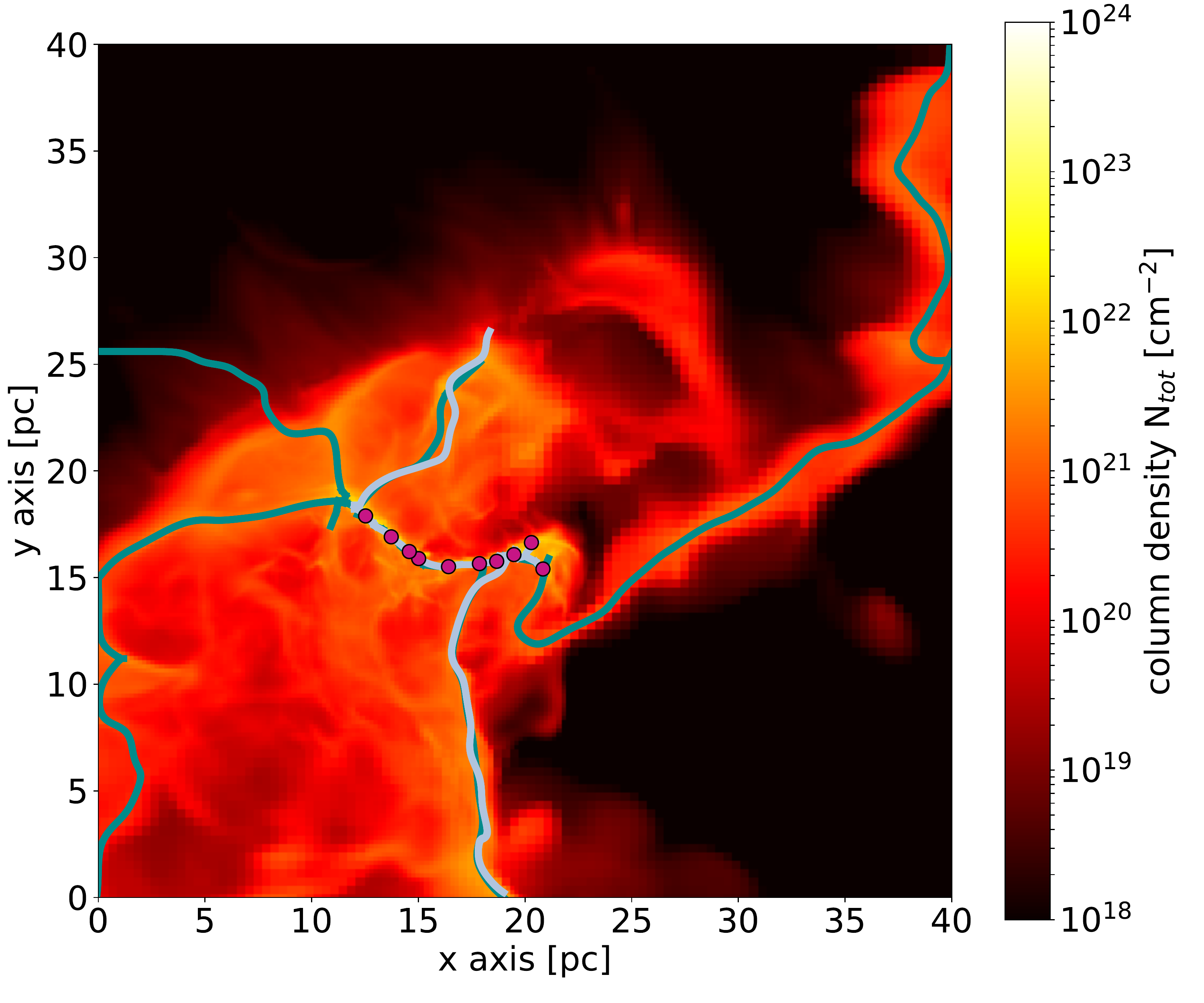}
    \caption{Column density map of \texttt{M3} at $t$ = 3.8 Myr when projected along the $z$-axis.
        The over-plotted lines illustrate the high-density filaments found by \texttt{DisPerSe} based on the column density map (\textit{teal lines}) and the volume density cube (projected, \textit{light blue lines}, same as in Fig.~\ref{pic_results_example_ntot_lowhigh}).
        Additionally, we plot the projected position of the identified fragments with purple dots.
    }
    \label{pic_results_example_fils_3d2d}
\end{figure}

Analogously to Sect.~\ref{results_3d_mean}, we use \texttt{DisPerSe} to identify 2D filaments within the column density maps.
For this purpose, we convert the number density thresholds used in 3D into column density thresholds by assuming a path length of 0.1 pc, namely $n_{\rm th} = 3 \times 10^{19}$~cm$^{-2}$ (corresponding to $n_{\rm th} = 100$~cm$^{-3}$) and $n_{\rm th}=10^{21}$~cm$^{-2}$ (corresponding to $n_{\rm th}=5,000$~cm$^{-3}$).
We emphasise that the skeletons of the 2D filaments are independently identified based on the column density distributions and not the projections of the 3D filament skeletons.
Figure~\ref{pic_results_example_fils_3d2d} shows an example of the structures obtained, and Table~\ref{tab_results_03_prop2d} summarises their properties.
We see that, similar to the structures detected in 3D, the 2D filaments' properties are influenced by the identification threshold, with the lengths of the filaments becoming shorter and line masses higher with higher thresholds.

\begin{table}
        \centering
        \begin{tabular}{l|cccc}
                Quantity & min & mean & median & max \\ \hline
                \multicolumn{5}{l}{N$_{\rm th}$~=~3~$\times$~10$^{19}$ cm$^{-2}$} \\ \hline
                Col.~density N$_{\rm tot}$ [10$^{21}$ cm$^{-2}$]        &  0.7    &  8.4  &  4.2  &  313          \\ 
                Gas temperature T$_{\rm gas}$ [K]                       & 11.1    & 56.7  & 56.4  &   97.8        \\ 
                Total length $\ell_{\rm tot}$ [pc]                              &  1.8    &  4.8  & 56.4  &   58.5        \\ 
                Total mass M$_{\rm tot}$ [M$_\odot$]                    &  0.7    & 46.0  & 17.0  & 1654          \\ 
                Line mass M$_{\rm lin,2D}$ [M$_\odot$ pc$^{-1}$]                &  0.4    &  9.6  &  5.0  &  321.1        \\ \hline 
                \multicolumn{5}{l}{N$_{\rm th}$~=~10$^{21}$ cm$^{-2}$} \\ \hline
                Col.~density N$_{\rm tot}$ [10$^{21}$ cm$^{-2}$]        &  0.6    & 13.4  &  6.0  &  295          \\ 
                Gas temperature T$_{\rm gas}$ [K]                       & 13.2    & 51.8  & 51.1  &   97.8        \\ 
                Total length $\ell_{\rm tot}$ [pc]                              &  1.8    &  4.3  &  2.5  &   35.3        \\ 
                Total mass M$_{\rm tot}$ [M$_\odot$]                    &  1.3    & 71.8  & 24.1  & 1735          \\ 
                Line mass M$_{\rm lin,2D}$ [M$_\odot$ pc$^{-1}$]                &  0.6    & 16.5  &  7.4  &  371          \\ \hline 
        \end{tabular}
        \caption{Summary of properties of the 2D filaments.}
        \label{tab_results_03_prop2d}
\end{table}

Comparing the structures detected in 3D and 2D, however, reveals a more significant difference.
As shown in the example in Fig.~\ref{pic_results_example_fils_3d2d}, we do find 2D counterparts for all 3D filaments, but there are 2D filaments that do not have matching 3D filaments identified with the same identification threshold.
This finding is a consequence of the projection:
structures with high volume densities typically have high column densities; conversely other structures with lower volume densities can appear denser in column densities, amplified by projection.

\begin{figure*}
        \centering
        \includegraphics[width=\textwidth]{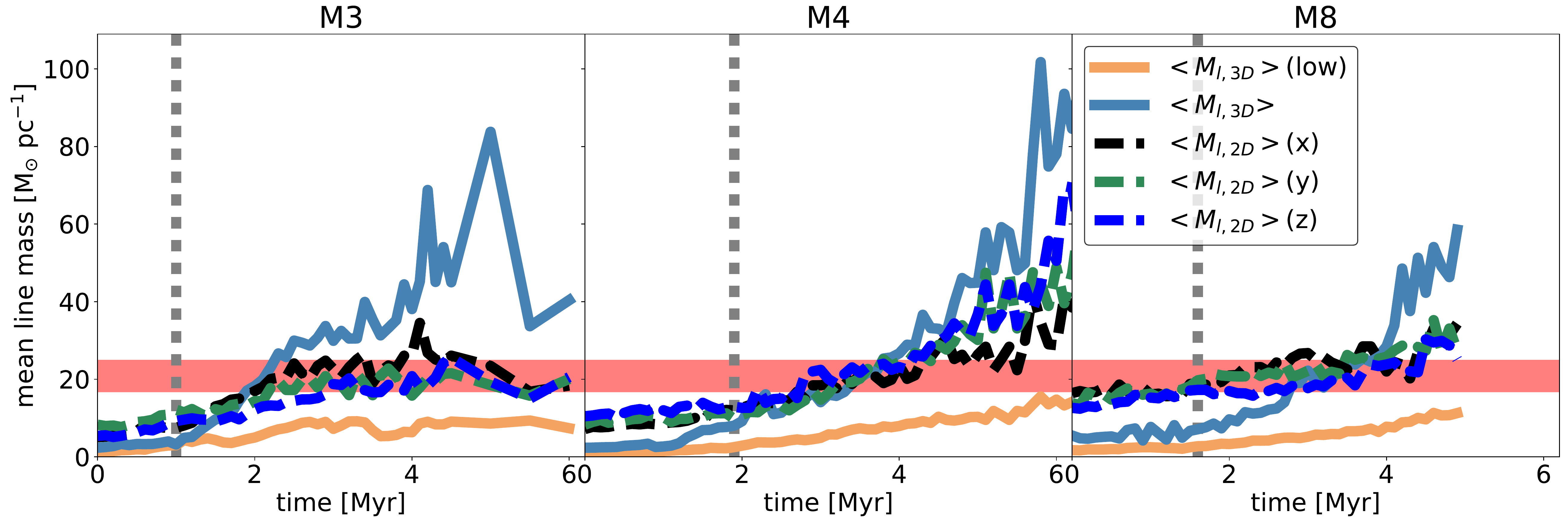}
        \caption{Evolution of $\langle M_{\rm lin,3D}\rangle$ (orange and blue lines) and $\langle M_{\rm lin, 2D}\rangle$ (black, green, and navy blue lines for projections along the $x$, $y$, and $z$ axes) for filaments that have been detected above $n_{\rm th}=5,000$~cm$^{-3}$ or $n_{\rm th} =10^{21}$~cm$^{-2}$, unless specified otherwise, in \texttt{M3} (\textit{left}), \texttt{M4} (\textit{middle}), and \texttt{M8} (\textit{right}).
      The red area marks the theoretical critical line mass for isothermal gas at 10--15~K.
      The grey dotted line shows the time step when the first fragment forms.
        }
        \label{pic_results_ml3dall}
\end{figure*}

This also influences the measured properties of the 2D filaments, such as the 2D line masses, $M_{\rm lin,2D}$.
Analogously to Fig.~\ref{pic_results_numfrags}, Fig.~\ref{pic_results_ml3dall} shows the value of $\langle {\rm M}_{\rm lin,2D} \rangle$ for the dense 2D filaments as a function of time. 
We see that the values lie between the values we measured based on the diffuse and dense 3D filaments.
If the dense 2D filaments exclusively represented the projection of dense 3D filaments, however, we would expect the trends in average line mass to evolve similarly.
This is not seen here.
On the contrary, the growth of $\langle {\rm M}_{\rm lin,2D} \rangle$ in the dense 2D filaments correlates with the growth of $\langle {\rm M}_{\rm lin,3D} \rangle$ of the diffuse 3D filaments.
As described before, this is because most of the dense 2D filaments are projections of diffuse 3D filaments, so the line mass of the 3D filaments is the main contributor of the average line mass of the dense 2D filaments.
Additional mass comes from material along the line of sight, so $M_{\rm lin,2D} > M_{\rm lin,3D}$.

In summary, for a given identification threshold, all 3D filaments have counterparts in column density maps, but not necessarily vice-versa.
Projection not only maps dense 3D filaments onto the plane of the sky, but also merges less dense structures along the same line of sight, producing structures that exceed the column density threshold.
Consequently, a comparison of the properties of 3D to 2D filaments is not directly possible if the corresponding identification thresholds are used for both samples.
However, if this is taken into account, the properties of 3D and 2D filaments evolve similarly, but with an offset due to the additional line of sight mass projected onto the 2D filaments.

The difference in the density distributions measured in 3D and 2D is even more obvious in the overall properties of the clouds.
We measure the DGMF using column density by defining clouds as coherent volumes of gas with minimal number densities of 100~cm$^{-3}$. 
We choose a minimum path length of 0.6~pc based on the assumption that filaments and fragments have radii of 0.3~pc.
We note that this path length is not identical to the path length we used for computing the identification threshold of the column density filaments, $n_{\rm th}$, before.
However, by using a minimum path length of 0.6~pc ensures that the fraction of gas we hereafter consider mirrors gas of the real 3D cloud.
This way, we cannot only more reliably compare the DGMFs measured based on the 2D data with those based on the 3D data, but also with DGMFs in observed molecular clouds.
As a results we obtain a minimal column density of $N_{\rm min}=2 \times 10^{20}$~cm$^{-2}$. 
We then compute
\begin{equation}
        {\rm DGMF}(t,N_{\rm dens})  = \frac{ \int_V N(t,\mathbf{x}) \, G(N_{\rm dens}) \,  dV}{\int_V N(t,\mathbf{x}) \, G({\rm 2 \times 10}^{20}\mathrm{~cm}^{-2}) \,  dV} \, ,
    \label{equ_results_dgmf_2D}
\end{equation}
with $N$ being the column density at time $t$ and within the pixel $\mathbf{x}$, $N_{\rm dens}$ the column density above which the gas is defined as dense and
\begin{equation}
        G(N_0) = \begin{cases}
                1 & N(t, \mathbf{x}) \geq N_0 \\
        0 & N(t, \mathbf{x}) < N_0
        \end{cases} .
    \label{equ_results_g_N}
\end{equation}
In Fig.~\ref{pic_results_fil3d_dgmf}, the dashed lines show the evolution of the DGMF measured using this equation.

Analogous to Sect.~\ref{results_3d_mean}, we use two thresholds for tracing the evolution of dense gas within the clouds, namely $N_{\rm dens}$~=~2~$\times$~10$^{21}$~cm$^{-2}$ (corresponding to $n_{\rm dens}$~=~1,000~cm$^{-3}$) and $N_{\rm dens}$~=~8~$\times$~10$^{21}$~cm$^{-2}$ (corresponding to $n_{\rm dens}$~=~5,000~cm$^{-3}$).
We see that for most of the time the DGMF is about an order of magnitude higher than the corresponding DGMFs calculated from the volume density distributions, but maximal values (70--80\% for $n_{\rm dens}$~=~1,000~cm$^{-3}$ and 30--50\% for $n_{\rm dens}$~=~5,000~cm$^{-3}$) agree with each other.
In the case of a lower value of $n_{\rm dens}$, we see that the 2D DGMFs are almost constant in time, or even slightly decreasing, in disagreement with
with the steady growth of the 3D DGMFs. 

In summary, we see that the DGMF measured in 2D deviates from the true 3D value for most of the initial evolution of a particular cloud.
The DGMF measured in column density may show a completely different temporal behaviour, particularly when a low column density threshold is used to define dense gas.
However, we also see that the individual filaments we identify in the 3D and 2D data and their properties agree decently with each other if the identification threshold in column density focuses on the range of volume density one wants to study and distinguishes unassociated gas along the line of sight.

\section{Summary \& Conclusions}\label{conclusions}

In this paper we analyse the properties and fragmentation of filaments forming within 3D AMR FLASH simulations of the self-gravitating, magnetised, supernova-driven ISM by \citet{IbanezMejia2016}. 
Our main results are as follows.

\begin{itemize}
        \item We find that the dense gas mass fraction (DGMF) steadily grows as a function of time. Although these clouds grow in mass as they accrete material from their environment, the DGMF continues growing in time, consistent with runaway gravitational collapse.
        \item We find that the average line masses of the filaments always increase in time, with significant differences depending on the volume (or column) density thresholds adopted for their identification. This and the continuously increasing filament-to-cloud mass ratio confirm that the gas of the parental clouds collapses into smaller scale structures, as the evolution of the DGMF has already indicated.
        \item Filaments already start to fragment well before their line masses reach the critical mass for the collapse of uniform density, self-gravitating, hydrostatic cylinders \citep{Ostriker1964b}. This is true both for the line masses of individual filaments, as well as for the the average of all identified filaments. This implies that the filaments in the simulation never resemble the isolated, hydrostatic configuration of \citet{Ostriker1964b} that is commonly used as the initial condition in analytic filament evolution and fragmentation models. Instead, they are embedded in the hierarchical collapse of the larger cloud, and thus subject to substantial surface pressures.
        \item We compare the performance of different filament finder codes (see Appendix \ref{a_filfinder}). We find that different codes clearly identify different structures, and further, that the filament properties derived depend strongly on the choice of input parameters. 
        \item We compare the properties of the filaments identified in 3D density distributions from the models with those identified in projected 2D column density distributions. We find that, for a given identification threshold density, all 3D filaments have counterparts in 2D column density data, but not vice versa. This is because the 2D filaments may also be composed of the overlap of more diffuse structures along the given line-of-sight that do not fulfil the identification criteria in 3D. As a consequence, the average properties of a sample of filaments and how they evolve in time are not well recovered from column density data. However, since all 3D filaments have counterparts in 2D, the correspondence is better in the case of individual isolated filaments.
\end{itemize}

Our results indicate that filament fragmentation is affected by the environment of the cloud they form in. 
In order to understand the onset and development of fragmentation, future theoretical studies likely need to abandon the hydrostatic initial condition and to consider the formation of filaments and their subsequent fragmentation together.

Furthermore, our results demonstrate that establishing common practices for how to define filaments in 3D and 2D data from simulations and observations is crucial for studying the properties and evolution of filaments and especially for comparing different filament studies with each other.
Studies using filament finders must thoroughly test applicability of the adopted algorithms to address the problem in question.

\begin{acknowledgements}
        The authors acknowledge the support ESO and its Studentship Programme provided.
        This research made use of astrodendro, a Python package to compute dendrograms of astronomical data (\url{http://www.dendrograms.org/}).
         This project has received funding from the European Union's Horizon 2020 research and innovation programme under grant agreement No 639459 (PROMISE). 
         JCI-M and M-MML received support from US NSF grant AST11-09395.  
         JCI-M was additionally supported by the DFG Priority Programme 157.
         M-MML also thanks the A. von Humboldt-Stiftung for support.
\end{acknowledgements}

        \bibliographystyle{aa} 
        \bibliography{ref}

\begin{thebibliography}{73}
\expandafter\ifx\csname natexlab\endcsname\relax\def\natexlab#1{#1}\fi

\bibitem[{{Abreu-Vicente} {et~al.}(2016){Abreu-Vicente}, {Ragan},
  {Kainulainen}, {Henning}, {Beuther}, \& {Johnston}}]{Abreu-Vicente2016}
{Abreu-Vicente}, J., {Ragan}, S., {Kainulainen}, J., {et~al.} 2016, ArXiv
  e-prints

\bibitem[{{Andr{\'e}} {et~al.}(2014){Andr{\'e}}, {Di Francesco},
  {Ward-Thompson}, {Inutsuka}, {Pudritz}, \& {Pineda}}]{Andre2014}
{Andr{\'e}}, P., {Di Francesco}, J., {Ward-Thompson}, D., {et~al.} 2014,
  Protostars and Planets VI, 27

\bibitem[{{Andr{\'e}} {et~al.}(2010){Andr{\'e}}, {Men'shchikov}, {Bontemps},
  {K{\"o}nyves}, {Motte}, {Schneider}, {Didelon}, {Minier}, {Saraceno},
  {Ward-Thompson}, {di Francesco}, {White}, {Molinari}, {Testi}, {Abergel}, \&
  {et al.}}]{Andre2010}
{Andr{\'e}}, P., {Men'shchikov}, A., {Bontemps}, S., {et~al.} 2010, \aap, 518,
  L102

\bibitem[{{Arzoumanian} {et~al.}(2011){Arzoumanian}, {Andr{\'e}}, {Didelon},
  {K{\"o}nyves}, {Schneider}, {Men'shchikov}, {Sousbie}, {Zavagno}, \& {et
  al.}}]{Arzoumanian2011}
{Arzoumanian}, D., {Andr{\'e}}, P., {Didelon}, P., {et~al.} 2011, \aua, 529, L6

\bibitem[{Bakes \& Tielens(1994)}]{Bakes1994TheHydrocarbons}
Bakes, E. L.~O. \& Tielens, A. G. G.~M. 1994, The Astrophysical Journal, 427,
  822

\bibitem[{{Ballesteros-Paredes}(2006)}]{Ballesteros-Paredes2006}
{Ballesteros-Paredes}, J. 2006, \mnras, 372, 443

\bibitem[{{Ballesteros-Paredes} \& {Mac Low}(2002)}]{Ballesteros2002}
{Ballesteros-Paredes}, J. \& {Mac Low}, M.-M. 2002, \apj, 570, 734

\bibitem[{{Barnard}(1927)}]{Barnard1927}
{Barnard}, E.~E. 1927, {Catalogue of 349 dark objects in the sky} (Chicago:
  University of Chicago Press)

\bibitem[{{Bergin} \& {Tafalla}(2007)}]{Bergin2007}
{Bergin}, E.~A. \& {Tafalla}, M. 2007, \araa, 45, 339

\bibitem[{{Bertoldi} \& {McKee}(1992)}]{Bertoldi1992}
{Bertoldi}, F. \& {McKee}, C.~F. 1992, \apj, 395, 140

\bibitem[{{Beuther} {et~al.}(2015){Beuther}, {Henning}, {Linz}, {Feng},
  {Ragan}, {Smith}, {Bihr}, {Sakai}, \& {Kuiper}}]{Beuther2015}
{Beuther}, H., {Henning}, T., {Linz}, H., {et~al.} 2015, \aap, 581, A119

\bibitem[{{Clarke} {et~al.}(2017){Clarke}, {Whitworth}, {Duarte-Cabral}, \&
  {Hubber}}]{Clarke2017}
{Clarke}, S.~D., {Whitworth}, A.~P., {Duarte-Cabral}, A., \& {Hubber}, D.~A.
  2017, \mnras, 468, 2489

\bibitem[{{Colombo} {et~al.}(2015){Colombo}, {Rosolowsky}, {Ginsburg},
  {Duarte-Cabral}, \& {Hughes}}]{SCIMES_Colombo2015}
{Colombo}, D., {Rosolowsky}, E., {Ginsburg}, A., {Duarte-Cabral}, A., \&
  {Hughes}, A. 2015, \mnras, 454, 2067

\bibitem[{{Contreras} {et~al.}(2016){Contreras}, {Garay}, {Rathborne}, \&
  {Sanhueza}}]{Contreras2016}
{Contreras}, Y., {Garay}, G., {Rathborne}, J.~M., \& {Sanhueza}, P. 2016,
  \mnras, 456, 2041

\bibitem[{Dalgarno \& McCray(1972)}]{Dalgarno1972HeatingRegions}
Dalgarno, A. \& McCray, R.~A. 1972, Annual Review of Astronomy and
  Astrophysics, 10, 375

\bibitem[{Dehnen \& Binney(1998)}]{Dehnen1998MassWay}
Dehnen, W. \& Binney, J. 1998, Monthly Notices of the Royal Astronomical
  Society, 294, 429

\bibitem[{{Enoch} {et~al.}(2006){Enoch}, {Young}, {Glenn}, {Evans}, {Golwala},
  {Sargent}, {Harvey}, {Aguirre}, {Goldin}, {Haig}, {Huard}, {Lange},
  {Laurent}, {Maloney}, {Mauskopf}, {Rossinot}, \& {Sayers}}]{Enoch2006}
{Enoch}, M.~L., {Young}, K.~E., {Glenn}, J., {et~al.} 2006, \apj, 638, 293

\bibitem[{{Federrath}(2016)}]{Federrath2016}
{Federrath}, C. 2016, \mnras, 457, 375

\bibitem[{{Fiege} \& {Pudritz}(2000{\natexlab{a}})}]{Fiege2000a}
{Fiege}, J.~D. \& {Pudritz}, R.~E. 2000{\natexlab{a}}, \mnras, 311, 85

\bibitem[{{Fiege} \& {Pudritz}(2000{\natexlab{b}})}]{Fiege2000b}
{Fiege}, J.~D. \& {Pudritz}, R.~E. 2000{\natexlab{b}}, \mnras, 311, 105

\bibitem[{{Fischera} \& {Martin}(2012)}]{Fischera2012a}
{Fischera}, J. \& {Martin}, P.~G. 2012, \aap, 542, A77

\bibitem[{{Fryxell} {et~al.}(2000){Fryxell}, {Olson}, {Ricker}, {Timmes},
  {Zingale}, {Lamb}, {MacNeice}, {Rosner}, {Truran}, \& {Tufo}}]{Fryxell2000}
{Fryxell}, B., {Olson}, K., {Ricker}, P., {et~al.} 2000, \apjs, 131, 273

\bibitem[{{G{\'o}mez} \& {V{\'a}zquez-Semadeni}(2014)}]{Gomez2014}
{G{\'o}mez}, G.~C. \& {V{\'a}zquez-Semadeni}, E. 2014, \apj, 791, 124

\bibitem[{{Gritschneder} {et~al.}(2017){Gritschneder}, {Heigl}, \&
  {Burkert}}]{Gritschneder2017}
{Gritschneder}, M., {Heigl}, S., \& {Burkert}, A. 2017, \apj, 834, 202

\bibitem[{{Gutermuth} {et~al.}(2009){Gutermuth}, {Megeath}, {Myers}, {Allen},
  {Pipher}, \& {Fazio}}]{Gutermuth2009}
{Gutermuth}, R.~A., {Megeath}, S.~T., {Myers}, P.~C., {et~al.} 2009, \apjs,
  184, 18

\bibitem[{{Hacar} \& {Tafalla}(2011)}]{Hacar2011}
{Hacar}, A. \& {Tafalla}, M. 2011, \aua, 533, A34

\bibitem[{{Hacar} {et~al.}(2013){Hacar}, {Tafalla}, {Kauffmann}, \&
  {Kov{\'a}cs}}]{Hacar2013}
{Hacar}, A., {Tafalla}, M., {Kauffmann}, J., \& {Kov{\'a}cs}, A. 2013, \aap,
  554, A55

\bibitem[{{Hartmann}(2002)}]{Hartmann2002}
{Hartmann}, L. 2002, \apj, 578, 914

\bibitem[{{Henshaw} {et~al.}(2016){Henshaw}, {Caselli}, {Fontani},
  {Jim{\'e}nez-Serra}, {Tan}, {Longmore}, {Pineda}, {Parker}, \&
  {Barnes}}]{Henshaw2016b}
{Henshaw}, J.~D., {Caselli}, P., {Fontani}, F., {et~al.} 2016, \mnras, 463, 146

\bibitem[{Hill {et~al.}(2012)Hill, Joung, Mac~Low, Benjamin, Matthew~Haffner,
  Klingenberg, \& Waagan}]{Hill2012}
Hill, A.~S., Joung, M. K.~R., Mac~Low, M.-M., {et~al.} 2012, The Astrophysical
  Journal, 750, 104

\bibitem[{{Ib{\'a}{\~n}ez-Mej{\'{\i}}a}
  {et~al.}(2016){Ib{\'a}{\~n}ez-Mej{\'{\i}}a}, {Mac Low}, {Klessen}, \&
  {Baczynski}}]{IbanezMejia2016}
{Ib{\'a}{\~n}ez-Mej{\'{\i}}a}, J.~C., {Mac Low}, M.-M., {Klessen}, R.~S., \&
  {Baczynski}, C. 2016, \apj, 824, 41

\bibitem[{{Ib{\'a}{\~n}ez-Mej{\'{\i}}a}
  {et~al.}(2017){Ib{\'a}{\~n}ez-Mej{\'{\i}}a}, {Mac Low}, {Klessen}, \&
  {Baczynski}}]{Ibanez-Mejia2017}
{Ib{\'a}{\~n}ez-Mej{\'{\i}}a}, J.~C., {Mac Low}, M.-M., {Klessen}, R.~S., \&
  {Baczynski}, C. 2017, \apj, subm., (ArXiv:1705.01779)

\bibitem[{{Inutsuka} \& {Miyama}(1992)}]{Inutsuka1992}
{Inutsuka}, S.-I. \& {Miyama}, S.~M. 1992, \apj, 388, 392

\bibitem[{{Jackson} {et~al.}(2010){Jackson}, {Finn}, {Chambers}, {Rathborne},
  \& {Simon}}]{Jackson2010}
{Jackson}, J.~M., {Finn}, S.~C., {Chambers}, E.~T., {Rathborne}, J.~M., \&
  {Simon}, R. 2010, \apjl, 719, L185

\bibitem[{{Juvela} {et~al.}(2012){Juvela}, {Malinen}, \&
  {Lunttila}}]{Juvela2012a}
{Juvela}, M., {Malinen}, J., \& {Lunttila}, T. 2012, \aap, 544, A141

\bibitem[{{Kainulainen} {et~al.}(2014){Kainulainen}, {Federrath}, \&
  {Henning}}]{Kainulainen2014}
{Kainulainen}, J., {Federrath}, C., \& {Henning}, T. 2014, Science, 344, 183

\bibitem[{{Kainulainen} {et~al.}(2013){Kainulainen}, {Ragan}, {Henning}, \&
  {Stutz}}]{Kainulainen2013c}
{Kainulainen}, J., {Ragan}, S.~E., {Henning}, T., \& {Stutz}, A. 2013, \aap,
  557, A120

\bibitem[{{Kainulainen} {et~al.}(2017){Kainulainen}, {Stutz}, {Stanke},
  {Abreu-Vicente}, {Beuther}, {Henning}, {Johnston}, \&
  {Megeath}}]{Kainulainen2017}
{Kainulainen}, J., {Stutz}, A.~M., {Stanke}, T., {et~al.} 2017, \aap, 600, A141

\bibitem[{{Koch} \& {Rosolowsky}(2015)}]{FilFinder_Koch2015}
{Koch}, E.~W. \& {Rosolowsky}, E.~W. 2015, \mnras, 452, 3435

\bibitem[{{K{\"o}nyves} {et~al.}(2015){K{\"o}nyves}, {Andr{\'e}},
  {Men'shchikov}, {Palmeirim}, {Arzoumanian}, {Schneider}, {Roy}, {Didelon},
  {Maury}, {Shimajiri}, {Di Francesco}, {Bontemps}, {Peretto}, {Benedettini},
  {Bernard}, {Elia}, {Griffin}, {Hill}, {Kirk}, {Ladjelate}, {Marsh}, {Martin},
  {Motte}, {Nguy{\^e}n Luong}, {Pezzuto}, {Roussel}, {Rygl}, {Sadavoy},
  {Schisano}, {Spinoglio}, {Ward-Thompson}, \& {White}}]{Koenyves2015}
{K{\"o}nyves}, V., {Andr{\'e}}, P., {Men'shchikov}, A., {et~al.} 2015, \aap,
  584, A91

\bibitem[{{Krumholz} \& {Tan}(2007)}]{Krumholz2007}
{Krumholz}, M.~R. \& {Tan}, J.~C. 2007, \apj, 654, 304

\bibitem[{{Larson}(1985)}]{Larson1985}
{Larson}, R.~B. 1985, \mnras, 214, 379

\bibitem[{{Lee} {et~al.}(1999){Lee}, {Myers}, \& {Tafalla}}]{Lee1999}
{Lee}, C.~W., {Myers}, P.~C., \& {Tafalla}, M. 1999, \apj, 526, 788

\bibitem[{{Li} {et~al.}(2016){Li}, {Urquhart}, {Leurini}, {Csengeri},
  {Wyrowski}, {Menten}, \& {Schuller}}]{Li2016}
{Li}, G.-X., {Urquhart}, J.~S., {Leurini}, S., {et~al.} 2016, \aap, 591, A5

\bibitem[{{Malinen} {et~al.}(2012){Malinen}, {Juvela}, {Rawlings},
  {Ward-Thompson}, {Palmeirim}, \& {Andr{\'e}}}]{Malinen2012}
{Malinen}, J., {Juvela}, M., {Rawlings}, M.~G., {et~al.} 2012, \aap, 544, A50

\bibitem[{{McKee} \& {Zweibel}(1992)}]{McKee1992}
{McKee}, C.~F. \& {Zweibel}, E.~G. 1992, \apj, 399, 551

\bibitem[{{Molinari} {et~al.}(2010){Molinari}, {Swinyard}, {Bally}, {Barlow},
  {Bernard}, {Martin}, {Moore}, {Noriega-Crespo}, {Plume}, {Testi}, {Zavagno},
  \& {Abergel}}]{Molinari2010}
{Molinari}, S., {Swinyard}, B., {Bally}, J., {et~al.} 2010, \pasp, 122, 314

\bibitem[{{Myers}(2011)}]{Myers2011}
{Myers}, P.~C. 2011, \apj, 735, 82

\bibitem[{{Nagasawa}(1987)}]{Nagasawa1987}
{Nagasawa}, M. 1987, Progress of Theoretical Physics, 77, 635

\bibitem[{Navarro {et~al.}(1996)Navarro, Frenk, \& White}]{Navarro1996TheHalos}
Navarro, J.~F., Frenk, C.~S., \& White, S. D.~M. 1996, The Astrophysical
  Journal, 462, 563

\bibitem[{Nešetřil {et~al.}(2001)Nešetřil, Milková, \&
  Nešetřilová}]{Nesetril2001}
Nešetřil, J., Milková, E., \& Nešetřilová, H. 2001, Discrete Mathematics,
  233, 3

\bibitem[{{Ostriker}(1964{\natexlab{a}})}]{Ostriker1964a}
{Ostriker}, J. 1964{\natexlab{a}}, \apj, 140, 1529

\bibitem[{{Ostriker}(1964{\natexlab{b}})}]{Ostriker1964b}
{Ostriker}, J. 1964{\natexlab{b}}, \apj, 140, 1056

\bibitem[{{Padoan} \& {Nordlund}(1999)}]{Padoan1999}
{Padoan}, P. \& {Nordlund}, {\AA}. 1999, \apj, 526, 279

\bibitem[{{Palmeirim} {et~al.}(2013){Palmeirim}, {Andr{\'e}}, {Kirk},
  {Ward-Thompson}, {Arzoumanian}, {K{\"o}nyves}, {Didelon}, {Schneider},
  {Benedettini}, {Bontemps}, {Di Francesco}, {Elia}, {Griffin}, {Hennemann},
  {Hill}, {Martin}, {Men'shchikov}, {Molinari}, {Motte}, {Nguyen Luong},
  {Nutter}, {Peretto}, {Pezzuto}, {Roy}, {Rygl}, {Spinoglio}, \&
  {White}}]{Palmeirim2013}
{Palmeirim}, P., {Andr{\'e}}, P., {Kirk}, J., {et~al.} 2013, \aap, 550, A38

\bibitem[{{Pineda} {et~al.}(2015){Pineda}, {Offner}, {Parker}, {Arce},
  {Goodman}, {Caselli}, {Fuller}, {Bourke}, \& {Corder}}]{Pineda2015}
{Pineda}, J.~E., {Offner}, S.~S.~R., {Parker}, R.~J., {et~al.} 2015, \nat, 518,
  213

\bibitem[{{Ragan} {et~al.}(2015){Ragan}, {Henning}, {Beuther}, {Linz}, \&
  {Zahorecz}}]{Ragan2015}
{Ragan}, S.~E., {Henning}, T., {Beuther}, H., {Linz}, H., \& {Zahorecz}, S.
  2015, \aap, 573, A119

\bibitem[{{Ragan} {et~al.}(2014){Ragan}, {Henning}, {Tackenberg}, {Beuther},
  {Johnston}, {Kainulainen}, \& {Linz}}]{Ragan2014}
{Ragan}, S.~E., {Henning}, T., {Tackenberg}, J., {et~al.} 2014, \aap, 568, A73

\bibitem[{{Rivera} {et~al.}(2015){Rivera}, {Loinard}, {Dzib}, {Ortiz-Le{\'o}n},
  {Rodr{\'{\i}}guez}, \& {Torres}}]{Rivera2015}
{Rivera}, J.~L., {Loinard}, L., {Dzib}, S.~A., {et~al.} 2015, \apj, 807, 119

\bibitem[{{Schmalzl} {et~al.}(2010){Schmalzl}, {Kainulainen}, {Quanz}, {Alves},
  {Goodman}, {Henning}, {Launhardt}, {Pineda}, \&
  {Rom{\'a}n-Z{\'u}{\~n}iga}}]{Schmalzl2010}
{Schmalzl}, M., {Kainulainen}, J., {Quanz}, S.~P., {et~al.} 2010, \apj, 725,
  1327

\bibitem[{{Schneider} {et~al.}(2012){Schneider}, {Csengeri}, {Hennemann},
  {Motte}, {Didelon}, {Federrath}, {Bontemps}, {Di Francesco}, {Arzoumanian},
  {Minier}, {Andr{\'e}}, {Hill}, {Zavagno}, {Nguyen-Luong}, {Attard},
  {Bernard}, {Elia}, {Fallscheer}, {Griffin}, {Kirk}, {Klessen}, {K{\"o}nyves},
  {Martin}, {Men'shchikov}, {Palmeirim}, {Peretto}, {Pestalozzi}, {Russeil},
  {Sadavoy}, {Sousbie}, {Testi}, {Tremblin}, {Ward-Thompson}, \&
  {White}}]{Schneider2012}
{Schneider}, N., {Csengeri}, T., {Hennemann}, M., {et~al.} 2012, \aap, 540, L11

\bibitem[{{Schneider} \& {Elmegreen}(1979)}]{Schneider1979}
{Schneider}, S. \& {Elmegreen}, B.~G. 1979, \apjs, 41, 87

\bibitem[{{Seifried} \& {Walch}(2015)}]{Seifried2015}
{Seifried}, D. \& {Walch}, S. 2015, \mnras, 452, 2410

\bibitem[{{Smith} {et~al.}(2014){Smith}, {Glover}, \& {Klessen}}]{Smith2014b}
{Smith}, R.~J., {Glover}, S.~C.~O., \& {Klessen}, R.~S. 2014, \mnras, 445, 2900

\bibitem[{{Smith} {et~al.}(2016){Smith}, {Glover}, {Klessen}, \&
  {Fuller}}]{Smith2016}
{Smith}, R.~J., {Glover}, S.~C.~O., {Klessen}, R.~S., \& {Fuller}, G.~A. 2016,
  \mnras, 455, 3640

\bibitem[{{Sousbie}(2011)}]{Sousbie2011a}
{Sousbie}, T. 2011, \mnras, 414, 350

\bibitem[{Sutherland \& Dopita(1993)}]{Sutherland1993CoolingPlasmas}
Sutherland, R.~S. \& Dopita, M.~A. 1993, The Astrophysical Journal Supplement
  Series, 88, 253

\bibitem[{Tammann {et~al.}(1994)Tammann, Loeffler, \&
  Schroeder}]{Tammann1994TheRate}
Tammann, G.~A., Loeffler, W., \& Schroeder, A. 1994, The Astrophysical Journal
  Supplement Series, 92, 487

\bibitem[{{Wang} {et~al.}(2016){Wang}, {Testi}, {Burkert}, {Walmsley},
  {Beuther}, \& {Henning}}]{Wang2016}
{Wang}, K., {Testi}, L., {Burkert}, A., {et~al.} 2016, \apjs, 226, 9

\bibitem[{{Wang} {et~al.}(2014){Wang}, {Zhang}, {Testi}, {van der Tak}, {Wu},
  {Zhang}, {Pillai}, {Wyrowski}, {Carey}, {Ragan}, \& {Henning}}]{Wang2014}
{Wang}, K., {Zhang}, Q., {Testi}, L., {et~al.} 2014, \mnras, 439, 3275

\bibitem[{{Zamora-Avil{\'e}s} {et~al.}(2017){Zamora-Avil{\'e}s},
  {Ballesteros-Paredes}, \& {Hartmann}}]{ZamoraAviles2017}
{Zamora-Avil{\'e}s}, M., {Ballesteros-Paredes}, J., \& {Hartmann}, L.~W. 2017,
  \mnras, 472, 647

\bibitem[{{Zhang} {et~al.}(2009){Zhang}, {Wang}, {Pillai}, \&
  {Rathborne}}]{Zhang2009}
{Zhang}, Q., {Wang}, Y., {Pillai}, T., \& {Rathborne}, J. 2009, \apj, 696, 268

\bibitem[{{Zucker} {et~al.}(2015){Zucker}, {Battersby}, \&
  {Goodman}}]{Zucker2015}
{Zucker}, C., {Battersby}, C., \& {Goodman}, A. 2015, \apj, 815, 23

\end{thebibliography}

        \appendix
\pagebreak

\section{Filament finders}\label{a_filfinder}

\subsection{Algorithms}\label{a_filfinder_algo}

There are a variety of algorithms publicly available for identifying filamentary structures in molecular clouds.
Naturally, there are more filament finders that work with 2D data than with 3D data, since 2D algorithms can directly be applied to observed data like intensity or column density maps, while 3D filament finders require data with a resolved third dimension, like the local standard of rest velocity observed spectroscopically.
The latter is not only observationally demanding, but computationally expensive to analyse.

In this Appendix, we present and compare the finder algorithms we have considered during our analysis and justify our choice of \texttt{DisPerSe} for the analysis performed in the main paper.

\subsubsection{\texttt{DisPerSe}}\label{a_filfinder_models_disperse}
\texttt{DisPerSe} \citep{Sousbie2011a} extracts coherent structures by evaluating the gradients between individual grid cells and the robustness of the topological features found. 
It was originally written for finding structures (both over- and under-densities) in cosmological data, but can be applied to other applications.

The advantage of \texttt{DisPerSe} is that it is independent of the content and dimension of data it receives.
Therefore, it can be directly applied to volume density cubes as well as column density maps and returns filamentary structures in both based on the same algorithms.

\subsubsection{\texttt{FilFinder}}\label{a_filfinder_models_filf}
\texttt{FilFinder} \citep{FilFinder_Koch2015} was written to extract filamentary structures in molecular clouds observed by the \textit{Herschel} Gould Belt Survey \citep{Andre2010}.
It does this by reducing the areas of interest (parts of molecular clouds with intensities above a specified threshold) to topological skeletons.
Therefore, each element of the skeletons represents the medial position of the areas of interest within the boundaries. 
Unfortunately, \texttt{FilFinder} can only be applied to 2D maps.

\subsubsection{\texttt{astrodendro}}\label{a_filfinder_models_astrodendro}
\texttt{astrodendro}\footnote{\url{http://dendrograms.readthedocs.io}} creates dendrogram trees representing the hierarchical structure of the underlying data.
That means that the code reveals how individual regions are connected with each other.
Those regions are then classified into trunks, branches, and leaves that represent molecular clouds, clumps and cores in our context.
Thus, \texttt{astrodendro} does not identify filamentary structures, but it is well suited for tracing fragments within the filaments or for preparing large data sets, so the actual filament finder can focus on the regions of interest.

\subsubsection{\texttt{SCIMES}}\label{a_filfinder_models_scimes}
\texttt{SCIMES} \citep{SCIMES_Colombo2015} works similarly to \texttt{astrodendro} (Sect.~\ref{a_filfinder_models_astrodendro}) using dendrograms.
However, while \texttt{astrodendro} simply detects structures, \texttt{SCIMES} weights the branches and leaves according to user-defined affinities (for example, minimal size, or maximal separation along position or velocity axes) and organises the dendrogram tree accordingly.
It returns weighted branches as clusters that, in our case, represent individual filaments.
We note that \texttt{SCIMES} itself is not a filament finder, but only returns regions that are likely to contain compact substructures, such as filaments.
In order to obtain the filaments one additionally needs a filament finder. 

There are some advantages to combining \texttt{SCIMES} and \texttt{DisPerSe}.
On the one hand, it is computationally more efficient to first create masks of the relevant regions before applying \texttt{DisPerSe} on those masks.
On the other hand, most filament finders return their structures without any weighting.
Hence, the users would need to distinguish the filaments from each other by hand, which is impractical for larger datasets like ours.
This step can be transferred to \texttt{SCIMES} as well, by applying the masks on the filament finder outputs.

\subsubsection{\texttt{Minimal Spanning Tree}}\label{a_filfinder_models_mst}
The minimal spanning trees \citep[MST, review by][]{Nesetril2001} algorithm is used for optimising costs by minimising the lengths of grids and efficiency of networks.
It can also be used to find coherent, filamentary structures as, for example, \citet{Wang2016} have demonstrated.
With the MST we can define filaments by connecting the leaves found by \texttt{astrodendro} representing pre-stellar cores in molecular clouds according to our requirements.

We use this method as an alternative method for identifying filaments in 3D.
\texttt{DisPerSe} also uses the MST algorithm to connect individual skeleton segments with each other.
The difference here is that we use MSTs for connecting the fragments we identified with \texttt{astrodendro} and define the straight lines between them as filaments.
Naturally, the separation between the fragments is on average much larger than those between \texttt{DisPerSe}'s segments, leading to less accurate curvatures.

\subsection{Comparison of identified filaments}\label{a_filfinder_results}

\begin{figure*}
        \centering
        \includegraphics[width=\textwidth]{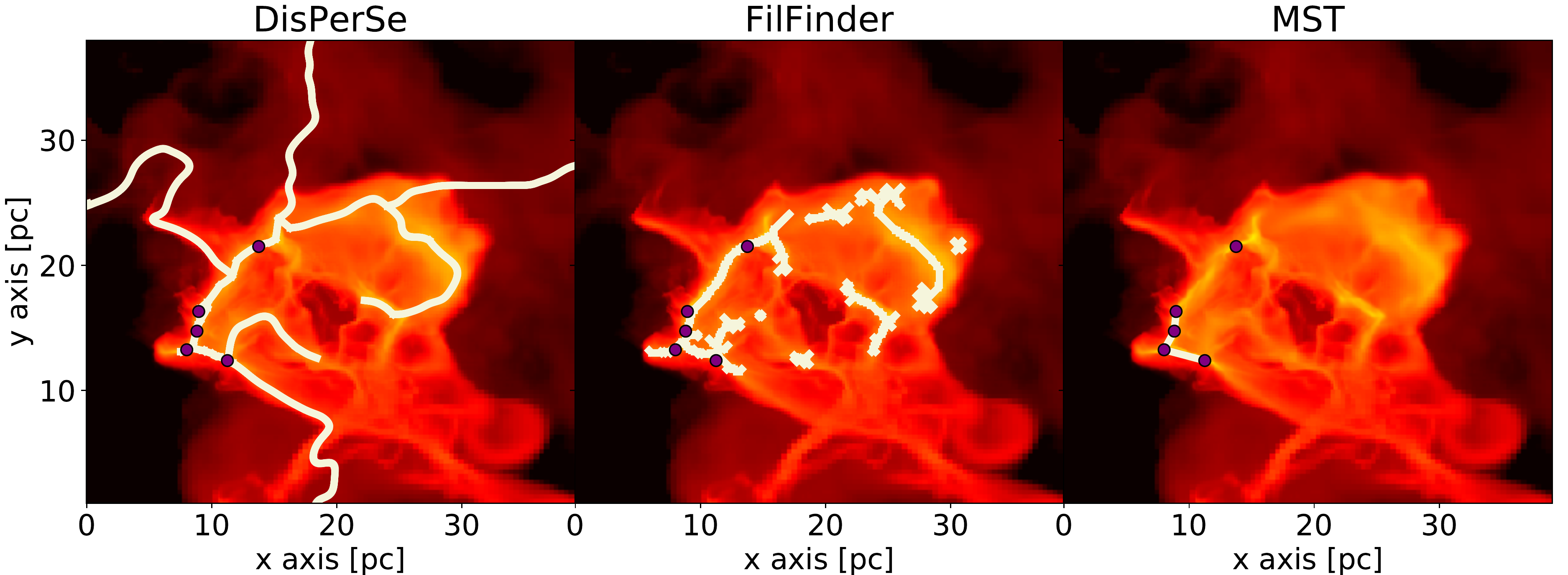}
        \caption{Example for structures identified by \texttt{DisPerSe} (\textit{left}), \texttt{FilFinder} (\textit{middle}), and \texttt{MST} applied to fragments from \texttt{astrodendro}  (\textit{right}).
                The background shows the column density map of \texttt{M4} at $t$ = 2.7 Myr along the z axis, ranging between 10$^{18}$ to 10$^{24}$ cm$^{-2}$.
                The purple dots represent the position of cores identified by \texttt{astrodendro}.
                The white lines illustrate filamentary structures found by the three filament finders, applying a column density cut at N$_{\rm th} = 1.5 \times 10^{21}$~cm$^{-2}$. 
        }
        \label{pic_appfilfinder_example_filfinders}
\end{figure*}

For evaluating and interpreting the results we have presented in Sect.~\ref{results} it is essential to understand and compare the performance of the underlying filament finder algorithms. 
Figure~\ref{pic_appfilfinder_example_filfinders} gives an example of this.
The background of each panel shows the column density map of the \texttt{M4} model projected along the $z$-axis.
The purple dots represent the position of the fragments identified by \texttt{astrodendro}.
The white lines illustrate the filaments found by \texttt{DisPerSe} (\textit{left}), \texttt{FilFinder} (\textit{center}), and \texttt{MST} (\textit{right}) applied to fragments from \texttt{astrodendro}, when using a column density threshold of N$_{\rm th}$~=~1.5~$\times$~10$^{21}$~cm$^{-2}$.
One sees that the codes return widely varying structures, although they agree well where the column density is highest.
Only \texttt{DisPerSe} follows the filaments that connect the cloud to the ISM.
This has a huge impact on the physical properties such as total length, enclosed mass and the field of interest in general.

\begin{figure*}
        \centering
        \includegraphics[width=\textwidth]{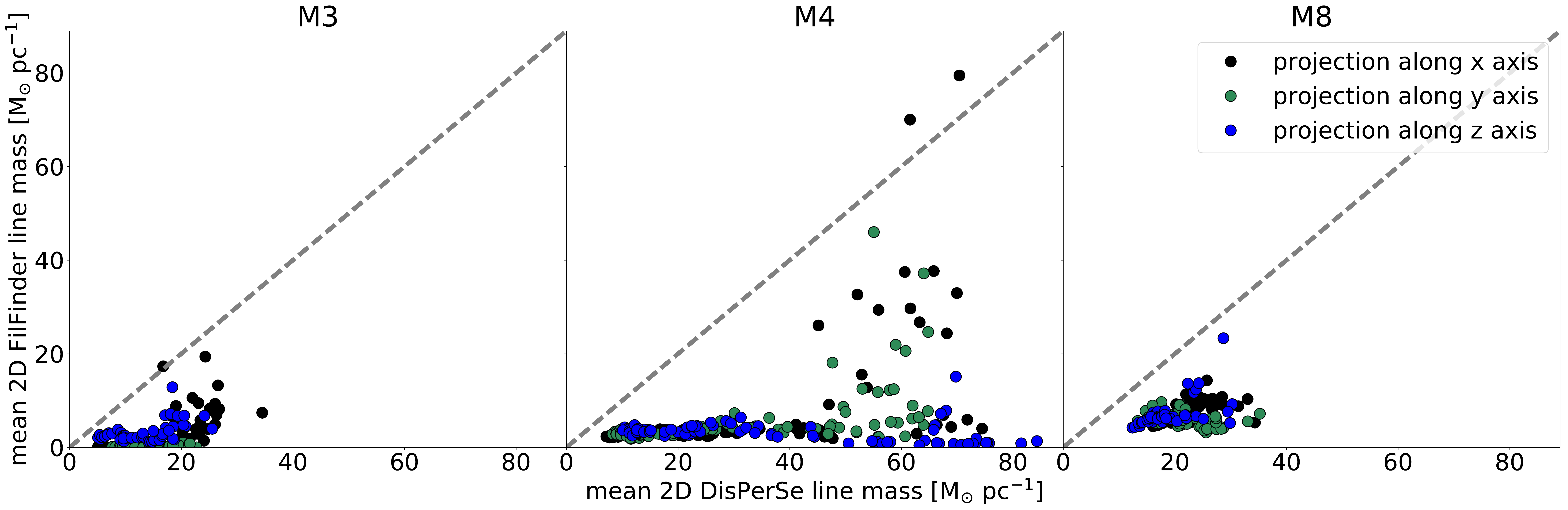}
        \caption{Comparison of mean 2D line masses, $\langle M_{lin,2D}\rangle$, derived with all filaments at the time steps of the simulations based on column density maps.
                The maps have been produced by projecting the volume density cubes along the $x$-axis (black dots), $y$-axis (green dots), and $z$-axis (blue dots).
                The values plotted correspond to the filaments identified by \texttt{DisPerSe} (ordinate) and \texttt{FilFinder} (abscissa).
        }
        \label{pic_appfilfinder_compml2d_disp_filf}
\end{figure*}

\begin{figure*}
        \centering
        \includegraphics[width=0.99\textwidth]{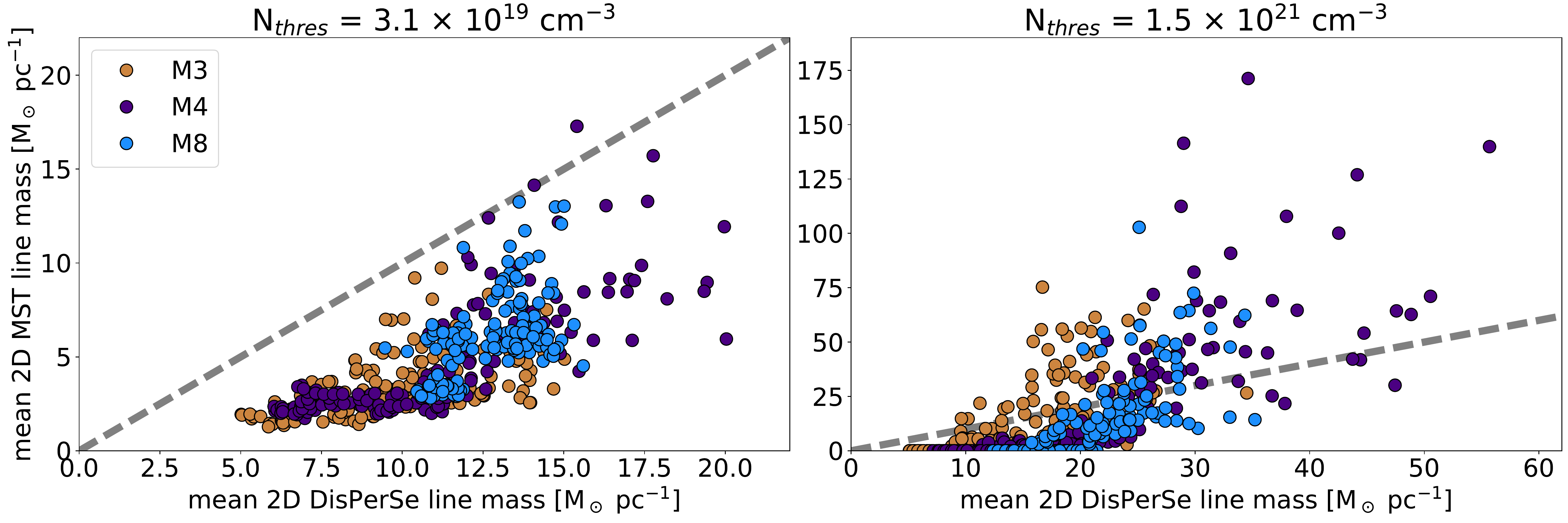}
        \caption{Comparison of $\langle M_{lin,2D}\rangle$ derived with all filaments derived using 2D column density maps at the time steps of the simulations.
                The values plotted correspond to the filaments identified by \texttt{DisPerSe} (ordinate) and by connecting \texttt{astrodendro} cores with \texttt{MST} (abscissa)
        }
        \label{pic_appfilfinder_compml2d_disp_mst}
\end{figure*}

\begin{figure*}
        \centering
        \includegraphics[width=\textwidth]{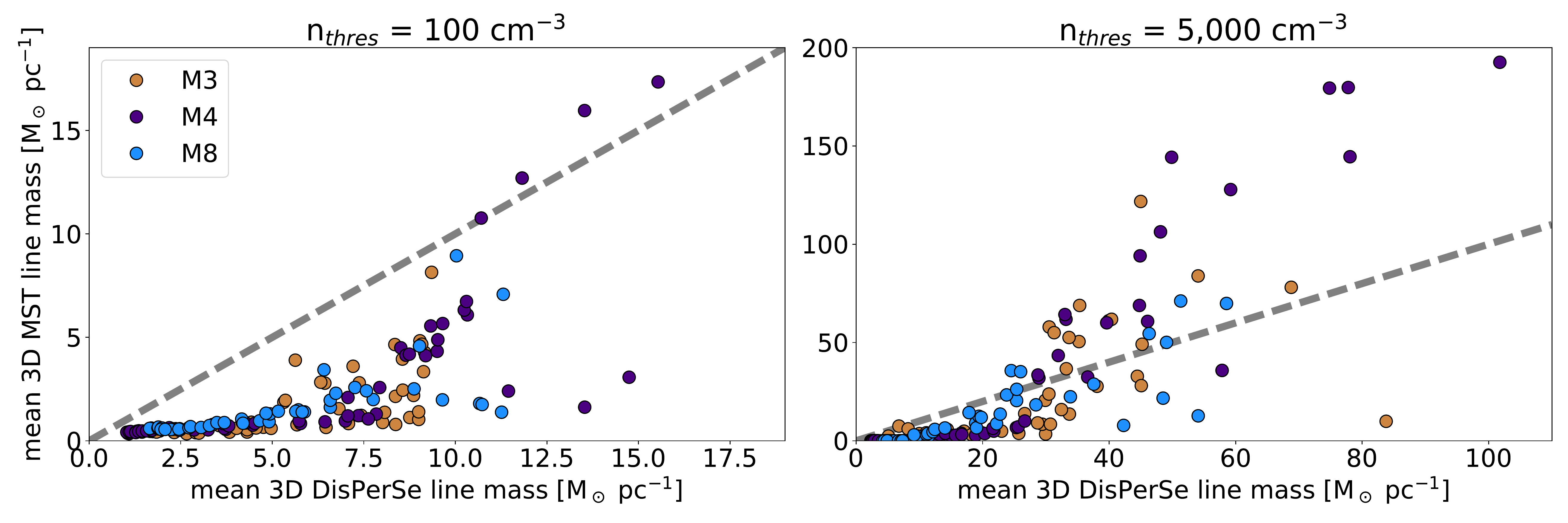}
        \caption{Comparison of $\langle M_{lin,3D}\rangle$ with all filaments derived using volume density cubes at the time steps of the simulations.
                The values plotted correspond to the filaments identified by \texttt{DisPerSe} (ordinate) and by connecting \texttt{astrodendro} cores with \texttt{MST} (abscissa).
        }
        \label{pic_appfilfinder_compml3d_disp_mst}
\end{figure*}

Since the different methods do not identify the same structures we cannot compare filaments individually. 
For evaluating the influence the underlying algorithms print on the structures, we measure the average line masses of all filaments detected within the respective cloud at a given time (analogously to Sect.~\ref{results_3d_mean}).
In Figs.~\ref{pic_appfilfinder_compml2d_disp_filf}, \ref{pic_appfilfinder_compml2d_disp_mst} and \ref{pic_appfilfinder_compml3d_disp_mst} we compare those average line masses based on the filaments returned by the individual codes in 2D and 3D, respectively.
We see that not only does the morphology of the filaments differ significantly using different codes, but also the properties of the structures.

We focus our further analysis on the filamentary skeletons identified by \texttt{DisPerSe} because we can automatically run the code on both our 2D and 3D and work with structures based on the same algorithm and parameter dependence.
Furthermore, \texttt{DisPerSe} is the code that is least sensitive to the input parameters since it is the only code that considers gradients in the matter distribution automatically, giving the skeletons a physical meaning.

\end{document}